%% file: CAI.tex
\documentclass[runningheads]{llncs}
\usepackage[T1]{fontenc}
\usepackage{graphicx}
\usepackage{tikz}

\usepackage{amsmath}
\usepackage{amssymb}

\usepackage{verbatim}
\usepackage[linesnumbered]{algorithm2e}
\usepackage[caption=false]{subfig}

\usepackage{tikz-qtree}

\spnewtheorem{claim*}[lemma]{Claim}{\itshape}{\normalfont}
\spnewtheorem*{cproof}{Proof of Claim}{\itshape}{\rmfamily}

\newenvironment{claimproof}
{\begin{cproof}
}
{\qed
\end{cproof}
}

\newenvironment{proof*}
{\begin{proof}
	}
	{\qed
\end{proof}}

\newcommand{\s}[1]{ \langle #1 \rangle}
\newcommand{\p}[1]{[\![ #1 ]\!]}
\newcommand{\angl}[1]{\langle #1 \rangle}

\newcommand{\ci}{\, \hat{\circ} \,}

\begin{document}
\title{How to decide Functionality of Compositions of Top-Down Tree Transducers} 
%
%
\author{Sebastian Maneth\inst{1} \and
	Helmut Seidl\inst{2} \and
	Martin Vu\inst{1}}
\authorrunning{S. Maneth,
	H. Seidl, and
	M. Vu}
%
\institute{Universit\"at Bremen, Germany \and
	TU M\"unchen, Germany
}

\maketitle

\begin{abstract}
We prove that functionality of compositions of top-down tree transducers is
decidable by reducing the problem to the functionality of one
top-down tree transducer with look-ahead.
\end{abstract}

\input{introduction}
\input{preliminaries}
\input{mainPartIntroduction}

\input{mainPartConstruction}
\input{CompositionNTransducers}

\input{conclusion}

\bibliographystyle{splncs04}
\bibliography{mybib}

\newpage
\appendix
\input{appendix}

\input{appendix-ii}
\input{appendix-iii}

\input{appendix-iv}
\end{document}

%% file: introduction.tex
\section{Introduction}\label{introduction}
Tree transducers are fundamental devices that were invented in the 1970's in
the context of compilers and mathematical linguistics.
Since then they have been applied in a huge variety of contexts
such as,
e.g., programming languages~\cite{DBLP:journals/lisp/MatsudaIN12}, 
security~\cite{DBLP:journals/iandc/KustersW07},
or XML databases~\cite{DBLP:conf/icde/HakutaMNI14}.

The perhaps most basic type of tree transducer is the
top-down tree transducer~\cite{DBLP:journals/jcss/Thatcher70,DBLP:journals/mst/Rounds70} 
(for short \emph{transducer}).
One important decision problem for transducers concerns \emph{functionality}:
given a (nondeterministic) transducer, does it realize a function?
This problem was shown to be decidable by {\'{E}sik}~\cite{DBLP:journals/actaC/Esik81}
(even in the presence of look-ahead);
note that this result also implies the decidability of equivalence of deterministic
transducers~\cite{DBLP:journals/actaC/Esik81}, see also~\cite{DBLP:journals/jcss/EngelfrietMS09,DBLP:journals/ijfcs/Maneth15}.

A natural and fundamental question is to ask whether functionality can also
be decided for \emph{compositions of transducers}.
It is well known that compositions of transducers form a proper hierarchy,
more precisely: compositions of $n+1$ transducers
are strictly more expressive than compositions of $n$~transducers~\cite{DBLP:journals/mst/Engelfriet82}.
Even though transducers are well studied, the question of deciding functionality for
compositions of transducers has remained open.
In this paper we fill this gap and show that the question can be answered affirmatively.

Deciding functionality for compositions of transducers has several applications.
For instance, if an arbitrary composition of (top-down and bottom-up) tree transducers
 is functional, then an equivalent 
deterministic transducer with look-ahead can be constructed~\cite{DBLP:journals/ipl/Engelfriet78}.
Together with our result this implies that it is decidable for such a composition
whether or not it is
definable by a deterministic transducer with look-ahead;
note that the construction of such a single deterministic transducer improves efficiency,
because it removes the need of computing intermediate results of the composition.
Also other recent definability results can now be generalized to compositions:
for instance, given such a composition we can now decide whether or not
an equivalent linear transducer or an equivalent homomorphism exists~\cite{DBLP:conf/dlt/ManethSV21} (and if so, construct it).

Let us now discuss the idea of our proof in detail.
Initially, we consider a composition $\tau$ of two transducers
$T_1$ and $T_2$. 
Given $\tau$,
we construct a `candidate' transducer with look-ahead $M$ 
with the property that $M$ is functional if and only if
$\tau$ is functional.
Our construction of $M$ is an extension of the product construction in~\cite[p.~195]{DBLP:journals/iandc/Baker79b}.
The latter constructs a transducer $N$ (\emph{without} look-ahead)
that is obtained by translating the right-hand sides of the rules of $T_1$ by the
transducer $T_2$. It is well-known that in general, the transducer $N$ is \emph{not} equivalent
to $\tau$~\cite{DBLP:journals/iandc/Baker79b} and thus $N$ may not be functional
even though $\tau$ is.
This is due to the fact that the transducer $T_2$ may 
\begin{itemize}
\item copy or
\item delete input subtrees.
\end{itemize}
Copying of an input tree means that the tree is translated several times and
in general by different states.
Deletion means that 
in a translation rule a particular input subtrees is not translated at all.

Imagine that $T_2$ copies and translates an input subtree in two different states $q_1$ and $q_2$,
so that the domains $D_1$ and $D_2$ of these states differ and moreover, $T_1$ nondeterministically produces
outputs in the union of $D_1$ and $D_2$.
Now the problem that arises in the product construction of $N$ is that $N$ needs to guess
the output of $T_1$, however, the two states corresponding to $q_1$ and $q_2$ \emph{cannot}
guarantee that the same guess is used. 
However, the same guess may be used.
This means that $N$ (seen as a binary relation) is a superset of $\tau$. 
To address this problem we show that it suffices to change $T_1$ so that it only outputs trees in
the intersection of $D_1$ and $D_2$. Roughly speaking this can be achieved by changing $T_1$ so
that it runs several tree automata in parallel, in order to carry out the necessary domain checks.

Imagine now a transducer $T_1$ that translates two input subtrees in states $q_1$ and
$q_2$, respectively, but has no rules for state $q_2$. This means that the translation of $T_1$
(and of $\tau$) is empty.
However, the transducer $T_2$ deletes the position of $q_2$. This causes the translation of $N$
to be non-empty. 
To address this problem we equip $N$ with look-ahead.
The look-ahead checks if the input tree is in the domains of all states
of $T_1$ translating the current input subtree.

Finally, we are able to generalize the result to arbitrary
compositions of transducers $T_1, \dots, T_n$.
For this, we apply the extended composition described above to the transducers
$T_{n-1}$ and $T_n$, giving us the transducer with look-ahead $M$.
The look-ahead of $M$ can be removed and  incorporated into the
transducer $T_{n-2}$ using a composition result of~\cite{DBLP:journals/iandc/Baker79b}. The resulting composition of $n-1$ transducers is
functional if and only if the original composition is. 

The details of all our proofs can be found in the Appendix.

%% file: preliminaries.tex
\section{Top-Down Tree Transducers}\label{top-down}
\label{sec:preliminaries}
For $k\in\mathbb{N}$, we denote by $[k]$ the set $\{1,\dots,k\}$.
Let $\Sigma=\{e_1^{k_1},\dots, e_n^{k_n}\}$ be a \emph{ranked alphabet}, where
$e_j^{k_j}$ means that the symbol $e_j$ has
\textit{rank} $k_j$.
By $\Sigma_k$ we denote the set of  all symbols of $\Sigma$ which have rank $k$.
The set $T_\Sigma$ of 
\textit{trees over $\Sigma$} 
consists of all strings of the form 
$a(t_1,\dots, t_k)$, where
$a\in \Sigma_k$, $k\geq 0$, and $t_1,\dots,t_k \in T_\Sigma$. 
Instead of $a()$ we simply write $a$.
We fix the set $X$ of \emph{variables} as $X=\{x_1,x_2,x_3,\dots \}$.

Let $B$ be an arbitrary set. We define
$T_\Sigma[B]=T_{\Sigma'}$ where $\Sigma'$ is obtained from $\Sigma$
by $\Sigma'_0=\Sigma_0\cup B$ while for all $k>0$, $\Sigma'_k =
\Sigma_k$. In the following, let $A,B$ be arbitrary sets.
We let $A(B)=\{a(b)\mid a\in A,b\in B\}$.

\begin{definition}
A \emph{top-down tree transducer}  $T$
(or \emph{transducer} for short) is a tuple of the form $T=(Q,\Sigma,\Delta,R, q_0)$ where
$Q$ is a finite set of \emph{states},
$\Sigma$ and $\Delta$ are the  \emph{input} and \emph{output ranked alphabets}, respectively, disjoint with $Q$,
$R$ is a \emph{finite set of rules}, and
$q_0\in Q$ is the \emph{initial state}.
The rules contained in $R$ are  of the form
$q(a(x_1,\dots,x_k)) \to t$, where
$q\in Q$, $a\in\Sigma_k$, $k\geq 0$ and
$t$ is a tree in $T_\Delta [Q(X)]$.
\end{definition}

If  $q(a(x_1,\dots,x_k)) \to t \in R$ then we call $t$ a right-hand side of $q$
and $a$.
The rules of $R$ are used as rewrite rules in the natural way, as illustrated by the following
example.

\begin{example}\label{introduction example}
	Consider the transducer $T=(\{q_0,q\}, \Sigma, \Delta, R, q_0)$
	where $\Sigma_0=\{e\}$, $\Sigma_1 =\{a\}$, $\Delta_0=\{e\}$, $\Delta_1=\{a\}$
	and $\Delta_2=\{f \}$
	and $R$ consists the following rules (numbered $1$ to $4$):
	\begin{center}
		$\begin{array}{lclclcl}
			1:\ q_0 (a(x_1))  & \rightarrow  & f (q(x_1), q_0(x_1)) & \quad & 2:\ q_0(e)  & \rightarrow  &  e\\
				3:\ q(a (x_1))  & \rightarrow  & a (q (x_1)) & \quad & 	4:\ q(e)  & \rightarrow  & e.  
		\end{array}$
	\end{center}
	On input $a(a(e))$, the transducer $T$ produces the output tree $f( a(e), f(e,e) )$
	as follows
	\begin{figure}[h]\label{translation figure}
		\begin{center}
		\begin{tikzpicture}
		\Tree [. $q_0$ [.\node (1) {$a$}; [.$a$ $e$ ] ] ]
		\draw[double,arrows=-stealth]   (0.3,-1.5)--(0.8,-1.5);
		\draw  (0.5,-2) node {$1$};
		\begin{scope}[xshift=1.4cm]
		\Tree  [. $f$ [.$q$ [.$a$ $e$ ] ] [. $q_0$ [.$a$ $e$ ] ] ] 
		\end{scope}
		\draw[double,arrows=-stealth]   (2.2,-1.5)--(2.7,-1.5);
		\draw  (2.4,-2) node {$1$};
		\begin{scope}[xshift=3.4cm]
		\Tree  [. $f$ [.$q$ [.$a$ $e$ ] ] [. $f$ [.$q$ $e$ ] [.$q_0$ $e$ ] ] ] 
		\end{scope}
		\draw[double,arrows=-stealth]   (4.3,-1.5)--(4.8,-1.5);
		\draw  (4.5,-2) node {$4$};
		\begin{scope}[xshift=5.4cm]
		\Tree  [. $f$ [.$q$ [.$a$ $e$ ] ] [. $f$ [. $e$ ] [.$q_0$ $e$ ] ] ] 
		\end{scope}
		\draw[double,arrows=-stealth]   (6.3,-1.5)--(6.8,-1.5);
		\draw  (6.5,-2) node {$3$};
		\begin{scope}[xshift=7.4cm]
		\Tree  [.$f$ [.$a$ [.$q$ $e$ ] ] [. $f$ [. $e$ ] [.$q_0$ $e$ ] ] ] 
		\end{scope}
		\draw[double,arrows=-stealth]   (8.3,-1.5)--(8.8,-1.5);
		\draw  (8.5,-2) node {$4$};
		\begin{scope}[xshift=9.4cm]
		\Tree  [. $f$ [.$a$  $e$ ] [. $f$ [. $e$ ] [.$q_0$ $e$ ] ] ] 
		\end{scope}
		\draw[double,arrows=-stealth]   (10.3,-1.5)--(10.8,-1.5);
		\draw  (10.5,-2) node {$2$};
		\begin{scope}[xshift=11.4cm, yshift=-0.5cm]
		\Tree  [. $f$ [.$a$  $e$ ] [. $f$ [. $e$ ] [. $e$ ] ] ] 
		\end{scope}
		\end{tikzpicture}
	\end{center}
	\end{figure}\newline
Informally, when processing a tree $s\in T_\Sigma$, the transducer $T$ produces a tree $t$ in which all proper subtrees of
$s$ occur as disjoint subtrees of $t$, `ordered' by size. 
As the reader may realize, given an input tree $s$ of size $n$, the transducer $T$ produces an output tree
that is of size $(n^2+n)/2$.
Hence, this translation has quadratic size increase, i.e.,
the size of the output tree is a most quadratic in size of the input tree.
Note that transducers can have polynomial or exponential size
increase~\cite{DBLP:journals/iandc/AhoU71}.  $\hfill\qed$
\end{example}

Let $s\in T_\Sigma$. Then $T(s)$ contains all trees in $T_\Delta$ obtainable from 
$q_0(s)$ by applying rules of $T$. 

Clearly, $T$ defines a binary relation over $T_\Sigma$
and $T_\Delta$. In the following, we denote by $\mathcal{R}(T)$ the binary relation that the transducer
$T$ defines. 
We say that the transducer $T$ is \emph{functional} if the relation $\mathcal{R}(T)$ is a function.
Let $q$ be a state of $T$. We denote by $\text{dom}(q)$ the \emph{domain} of $q$, 
i.e., the set of all trees $s \in T_\Sigma$
for which some tree $t\in T_\Delta$ is obtainable from $q(s)$ by applying rules
of $T$.
We define the  \emph{domain} of $T$ by  $\text{dom}(T)=\text{dom}(q_0)$.   
For instance in Example~\ref{introduction example}, $\text{dom}(T)=T_\Sigma$.
However, if we remove the rule $1$ for instance then the domain of $T$ shrinks to 
the set $\{e \}$.
We define $\text{dom} (q)$, the \emph{domain} of a state $q$ of $T$, analogously.

A transducer $T=(Q,\Sigma, \Delta, R, q)$ is a \emph{top-down tree automaton} (for short \emph{automaton})
if $\Sigma=\Delta$ and all rules of $T$ are of the form
$q(a(x_1,\dots,x_k))\rightarrow a(q_1(x_1),\dots q_k(x_k))$ where $a\in \Sigma_k$, $k\geq 0$.

 Let $T_1$ and $T_2$ be transducers.
 As $\mathcal{R}(T_1)$ and $\mathcal{R}(T_2)$ are relations, they can be composed.  
 Hence,
 \[
  \mathcal{R}(T_1)\circ \mathcal{R}(T_2)= \{(s,u) \mid \text{for some } t, (s,t)\in \mathcal{R}(T_1) \text{ and } (t,u)\in \mathcal{R}(T_2)\}.
 \]
 If  the output alphabet of  $T_1$ and the input alphabet of 
 $T_2$ coincide then the transducers $T_1$ and $T_2$  can be composed as well.
 The composition  $T_1 \ci T_2$ of the transducers $T_1$ and $T_2$ defines a tree translation
 as follows.
 On input $s$, the tree $s$ is first translated by $T_1$.
 Afterwards, the tree produced by $T_1$ is translated by $T_2$ which yields the output tree.
 Clearly, $T_1\ci T_2$ computes the relation $\mathcal{R} (T_1)\circ \mathcal{R} (T_2)$.
 We say that the composition $T_1\ci T_2$  is \emph{functional} if the relation
 $\mathcal{R} (T_1)\circ \mathcal{R} (T_2)$ is a function.

%% file: mainPartIntroduction.tex
\section{Functionality of Two-Fold Compositions}\label{mainpartintro}
In this section we show that for  a composition $\tau$ of two transducers, a transducer $M$
with look-ahead can be constructed such that $M$ is functional if and only if
$\tau$ is functional.
Before formally introducing the construction for $M$
and proving its correctness, 
we explain how to solve the challenges described in Section~\ref{introduction}, i.e.,
we show how to handle copying and deleting rules.
In the following, we call the product construction in~\cite[p.~195]{DBLP:journals/iandc/Baker79b}
simply the \emph{p-construction}.

To see how precisely we handle copying rules, consider the transducers $T_1$ and $T_2$.
Let the transducer $T_1$ consist of the rules
\begin{center}
	$\begin{array}{lclclcl}
	q_1(a(x_1)) &\rightarrow & b(q_1 (x_1)) & \quad &  q_1(e) &\rightarrow & e_i \mid i=1,2,3 \\
	\end{array}$
\end{center}
while transducer $T_2$ consist of the rules
\begin{center}
	$\begin{array}{lclclcl}
	q_2(b(x_1)) &\rightarrow & f(q'_2 (x_1), q_2'' (x_1)) & \quad &  q_2'(e_j) &\rightarrow & e \mid j=1,2\\
	q_2''(e_3) &\rightarrow & e'& \quad & q_2''(e_j) &\rightarrow & e \mid j=1,2.\\
	\end{array}$
\end{center}
The composition $\tau=T_1\ci T_2$ defines a relation that only contains a single pair:
$\tau$ only translates the tree $a(e)$ into $f(e,e)$.
Therefore, $\tau$ is functional.
For $T_1$ and $T_2$, the p-construction yields the transducer
$N$ with the rules
\begin{center}
	$\begin{array}{lclclcl}
	(q_1,q_2) (a(x_1)) &\rightarrow & f((q_1,q'_2) (x_1), (q_1,q_2'') (x_1)) & \quad &  (q_1,q_2')(e) &\rightarrow & e\\
	(q_1,q_2'')(e) &\rightarrow & e'& \quad & (q_1,q_2'')(e) &\rightarrow & e.\\
	\end{array}$
\end{center}
On input $a(e)$, 
the transducer $N$ can produce either $f(e,e)$ or $f(e,e')$.
Therefore, $N$ and  $\tau$ are clearly not equivalent.
Furthermore, the transducer $N$ is obviously not functional even though the composition $\tau$ is.

In order to obtain a better understanding of why this phenomenon occurs,
we analyze the behavior of $N$ and $\tau$ on input $a(e)$ in the following.

In the translation of $\tau$, 
the states $q_2'$ and $q_2''$ process the same
tree produced by $q_1$ on input $e$ due to the copying rule 
$q_2(b(x_1)) \rightarrow f(q'_2 (x_1), q_2'' (x_1))$. Furthermore,  $q_2'$ and $q_2''$ process a tree 
in $\text{dom} (q_2') \cap \text{dom} (q_2'')$. More precisely, $q_2'$ and $q_2''$  both
process either $e_1$ or $e_2$.

In the translation of $N$ on the other hand, due to the rule 
$(q_1,q_2) (a(x_1)) \rightarrow  f((q_1,q'_2) (x_1), (q_1,q_2'') (x_1))$,
the states $(q_1,q_2')$ and $(q_1,q_2'')$  process $e$ by `guessing independently' from each other 
what $q_1$ might have produced on input $e$. 
In particular, the problem is that $(q_1,q_2'')$ can apply the
rule $(q_1,q_2'') (e) \rightarrow e'$ which eventually leads to the production of  $f(e,e')$.
Applying this rule means that $(q_1,q_2'')$ 
guesses that $e_3$ is produced by $q_1$. 
While this guess is valid, i.e., $e_3$ is producible by $q_1$ on input
$e$, quite clearly $e_3\notin \text{dom}(q_2')$.

In general, guesses performed by states of $N$ cannot be `synchronized', i.e., 
we cannot guarantee that states guess the same tree.
Our solution to fix this  issue is to restrict
$(q_1,q_2')$ and $(q_1,q_2'')$ such that either state is only allowed to guess trees in 
$\text{dom} (q_2') \cap \text{dom} (q_2'')$. To understand why this approach works
in  general consider the following example. 

\begin{example}\label{plausible}
Let ${T}_1$ and ${T}_2$ be arbitrary transducers. Let ${\tau}={T}_1\ci {T}_2$ be functional.
Let ${T}_1$ on input $s$ produce either $b(t_1)$ or $b(t_2)$. Let ${T}_2$ contain
the rule  \[q_2(b(x_1))\rightarrow f(q_2^1(x_1), q_2^2(x_1))\] where $q_2$ is the 
initial state of ${T}_2$. The application of this rule 
effectively means that the states $q_2^1$ and $q_2^2$ process the same subtree produced by ${T}_1$.
Let $t_1,t_2\in  \text{dom} (q_2^1) \cap \text{dom} (q_2^2)$.
Informally speaking, it does not matter whether the state $q_2^1$
processes $t_1$ or $t_2$; for either input $q_2^1$  produces the same output tree $r$
and nothing else, otherwise, the functionality of $\tau$ is contradicted.
The same holds for $q_2^2$. $\hfill\qed$
\end{example}

Informally, Example~\ref{plausible} suggests that if
$(q_1,q_2'')$ and $(q_1,q_2'')$ only guess trees in 
$\text{dom} (q_2') \cap \text{dom} (q_2'')$, then it does not matter which tree exactly those states
guess if the composition is functional. The final result in either case is the same.
Quite clearly this is the case in our example. (In effect, $q_2''$ is forbidden to guess~$e_3$.)
Thus, restricting $(q_1,q_2')$ and $(q_1,q_2'')$
basically achieves the same result as synchronizing their guesses if the composition is functional.

Now the question is how exactly do we restrict the states of $N$?
Consider the states $(q_1,q_2')$ and $(q_1,q_2'')$ of $N$ in our example.
The trick is to restrict $q_1$ such that $q_1$ can only produce trees in $\text{dom} (q_2') \cap \text{dom} (q_2'')$. Thus any guess is guaranteed to be in $\text{dom} (q_2') \cap \text{dom} (q_2'')$.
In order to restrict which output trees $T_1$ can produce,
we compose $T_1$ with the \emph{domain automaton} of $T_2$. 

For an arbitrary transducer $T=(Q,\Sigma,\Delta, R,q)$,
the \emph{domain automaton} $A$ of $T$ is constructed 
analogous  to the automaton in \cite[Theorem~3.1]{DBLP:journals/mst/Engelfriet77}.
The  set of states of $A$ is the power set of ${Q}$
where $\{ q \}$
is the initial state of $A$.
The idea is that if in a translation of $T$ on input $s$, 
the states $q_1\dots,q_n$ process the node $v$ of $s$
then $\{q_1\dots,q_n\}$ processes the node $v$ of $s$ in a computation of $A$.
The rules of $A$ are thus defined as follows.

Let $S=\{q_1\dots,q_n\}$, $n> 0$, and $a\in \Sigma_k$.
In the following,  we denote  by  $\text{rhs}_T(q_j,a)$, where $j\in [n]$, the set of all right-hand sides of $q_j$ and $a$.
For all non-empty subsets $\Gamma_1\subseteq \text{rhs}_T(q_1,a),\dots, \Gamma_n\subseteq \text{rhs}_T(q_n,a)$,
we define a rule
\[
S (a(x_1,\dots,x_k))\rightarrow a(S_1 (x_1),\dots,S_k (x_k))
\]
where 
for $i\in [k]$, $S_i$ is defined as the set $\bigcup_{j = 1}^{n} \Gamma_j \s{x_i}$.
We denote by $\Gamma_j \s{x_i}$ the set of all states $q'$ such that
$q'(x_i)$ occurs in some tree $\gamma$ in $\Gamma_j$; e.g., for
\[
\Gamma_j=\{a(q(x_1), q'(x_2)),\ a(a(q_1(x_1), q_2 (x_2)), q_3(x_1))\},
\]
we have
$\Gamma_j \s{x_1} =\{q,q_1,q_3\}$ and $\Gamma_j \s{x_2} =\{q',q_2\}$.
We define that the state $\emptyset$ of $A$ realizes the identity.
Hence, the rules for the state $\emptyset$ are defined in the obvious way.

We now explain why
subsets $\Gamma_j$ of right-hand sides are used for the construction of rules of $A$. 
Recall that the idea is that if in a translation of $T$ on input $s$, 
the states $q_1\dots,q_n$ process the node $v$ of $s$
then $\{q_1\dots,q_n\}$ processes the node $v$ of $s$ in a computation of $A$.
Due to copying rules, multiple instances of a state $q_1$ may access $v$.
Two instance of $q_1$ may process $v$ in different manners.
This necessitates the use of subsets $\Gamma_j$ of right-hand sides.
For a better understanding, consider the following example.

\begin{example}
Let $T=(\{q_0,q\}, \Sigma, \Delta, R, q_0)$
where $\Sigma_0=\Delta_0=\{e\}$, $\Sigma_1 = \Delta_1=\{a\}$ and $\Sigma_2=\Delta_2=\{f\}$.
The set $R$ contains the following rules:
\begin{center}
	$\begin{array}{lclclcl}
	q_0 (a(x_1))  & \rightarrow  & f (q_0(x_1), q_0(x_1)) & \quad & q(a(x_1))  & \rightarrow  & e'\\ 
	q_0 (f (x_1, x_2))  & \rightarrow  & q_0 (x_1) & \quad &  q(f (x_1,x_2)) &\rightarrow  & e'\\
	 q_0 (f (x_1, x_2))  & \rightarrow  & f (q (x_1), q (x_2)) & \quad & q(e)  & \rightarrow  &  e'\\
	 q_0(e)  & \rightarrow  & e.\\
	\end{array}$
\end{center}
Consider the input tree $s=a(f(e,e))$. Clearly, on input $s$, the tree $f(e, f(e',e'))$
is producible by $T$. In this translation, two instances of the state $q_0$ process the
subtree $f(e,e)$ of $s$, however the instances of $q_0$ do not process $f(e,e)$
in the same way. The first instance of $q_0$ produces $e$ on input $f(e,e)$
while the second instance produces $f(e',e')$. These translations mean that
the states $q_0$ and $q$ process the leftmost $e$ of $s$.

Consider the domain automaton $A$ of $T$.
By definition, $A$  contains the rule
$\{q_0\} (a(x_1)) \rightarrow a (\{q_0\} (x_1))$ which is obtained from the
right-hand side of the rule $q_0 (a(x_1)) \rightarrow  f (q_0(x_1), q_0(x_1))$ of
$T$. To simulate that the states $q_0$ and $q$ process the leftmost $e$ of $s$
in the translation from $s$ to $f(e, f(e',e'))$, we clearly require the
rule
$\{q_0\} (f(x_1,x_2)) \rightarrow f (\{q_0,q\} (x_1) , \{q\} (x_2))$
obtained from the right-hand sides of the rules
$q_0 (f (x_1, x_2)  \rightarrow  q_0 (x_1)$ and 
$q_0 (f (x_1, x_2)   \rightarrow  f (q (x_1), q (x_2))$ of $T$.

For completeness, we list the remaining rules of $A$.
The automaton $A$ also contains the rules
\begin{center}
	$\begin{array}{rlclcrlcl}
\{q_0\} & (f(x_1,x_2)) & \rightarrow  & f (\{q\} (x_1) , \{q\} (x_2))        &\quad&  \{ q\} & (a (x_1)) & \rightarrow & a( \emptyset (x_1))\\
\{q_0\}  &(f(x_1,x_2)) & \rightarrow  & f (\{q_0\} (x_1) , \emptyset (x_2)) &\quad&  \{ q\} & (f (x_1,x_2)) & \rightarrow & f( \emptyset (x_1),  \emptyset (x_2))\\
\{q_0\} &(e) & \rightarrow  & e                                             &\quad & \{q\} & (e) & \rightarrow  & e.\\
  \emptyset &  (f (x_1,x_2)) & \rightarrow & f( \emptyset (x_1),  \emptyset (x_2))& \quad &\emptyset& (a (x_1)) & \rightarrow & a( \emptyset (x_1))     \\
 \emptyset & (e) & \rightarrow  & e.\\
	\end{array}$
\end{center}
For the rules of the state $\{q_0,q\}$  consider the following.
The right-hand sides of rules of $\{q_0,q\}$ are identical to the
right-hand sides of rules of $\{q_0\}$, i.e.,
 the rules  for $\{q_0,q\}$ are obtained  by substituting
$\{q_0 \}$  on the left-hand-side of rules of $A$ by $\{q_0 ,q\}$. $\hfill\qed$
\end{example}

\noindent
The automaton $A$ has the following property.
\begin{lemma}\label{Property of A}
	Let $S\neq \emptyset$ be a state of $A$.
	Then $s\in \text{dom} (S)$ if and only if $s\in \bigcap_{q\in S} \text{dom} (q)$.
\end{lemma}
\noindent
Obviously, Lemma~\ref{Property of A} implies that $A$ recognizes the domain of $T$.

Using the domain automaton $A$ of $T_2$, we transform $T_1$ into the transducer $\hat{T}_1$.
Formally, the transducer $\hat{T}_1$ is obtained  from $T_1$ and  $A$ using the p-construction. 
In our example, the transducer $\hat{T}_1$ obtained from $T_1$  and $T_2$ includes the following rules
\begin{center}
	$\begin{array}{rlcl}
	(q_1,\{q_2\}) & (a(x_1)) &\rightarrow & b( (q_1,\{q_2', q_2''\}) (x_1)) \\
	(q_1,\{q_2', q_2''\}) & (e) &\rightarrow & e_j  \\
	\end{array}$
\end{center}
where $j=1,2$. The state $(q_1,\{q_2\})$ is the initial state of $\hat{T}_1$.
Informally, the idea is that in a translation of $\hat{\tau}=\hat{T}_1\ci T_2$,
a tree produced by a state $(q,S)$ of $\hat{T}_1$ is only processed by states in $S$.
The following result complements this idea.
\begin{lemma}\label{Property of hatT1}
	If the state $(q,S)$ of $\hat{T}_1$ produces the tree $t$ and $S\neq\emptyset$ then $t\in  \bigcap_{q_2\in S} \text{dom}(q_2)$.
\end{lemma}
We remark that if 
a state of the form $(q,\emptyset)$ occurs
then it means that
in a translation of $\hat{\tau}$, no state of $T_2$ will process a tree produced by $(q,\emptyset)$. 
Note that as $A$ is nondeleting and linear, $\hat{T}_1$ defines the same relation as $T_1\ci A$~\cite[Th.~1]{DBLP:journals/iandc/Baker79b}.
Informally, the transducer $\hat{T}_1$ is a restriction of the transducer $T_1$
such that $\text{range} (\hat{T}_1)=\text{range} (T_1) \cap \text{dom} (T_2)$.
Therefore, the following holds.

\begin{lemma}\label{T1' property 4}
	$\mathcal{R} (T_1)\circ \mathcal{R} (T_2) =\mathcal{R}(\hat{T}_1)\circ \mathcal{R} (T_2)$.
\end{lemma}
Due to Lemma~\ref{T1' property 4}, we focus on $\hat{T}_1$ instead of $T_1$ in the following.

Consider the transducer $\hat{N}$ obtained from $\hat{T}_1$
and $T_2$ using the p-construction.
By construction, the states of $\hat{N}$  are of the form $((q,S),q')$ where $(q,S)$ is a state of $\hat{T}_1$
and $q'$ is a state of $T_2$. In the following, we write $(q,S,q')$ instead for better readability.
Informally, the state $(q,S,q')$ implies that in a translation of 
$\hat{\tau}$ the state $q'$ is supposed to process a tree
produced by $(q,S)$. Because trees produced by $(q,S)$ are only supposed to be processed by states in $S$, 
we only consider states $(q,S,q')$ where $q'\in S$.
For $\hat{T}_1$ and $T_2$, we obtain the transducer $\hat{N}$ with the following rules
	\begin{center}
		$\begin{array}{rlcl}
		(q_1,\{q_2\},q_2) & (a(x_1)) &\rightarrow &  f( (q_1,S,q_2') (x_1), (q_1,S,q_2'')(x_1) )  \\
		(q_1,S,q_2') & (e) &\rightarrow & e\\ 
		(q_1,S, q_2'') &(e) & \rightarrow & e\\
		\end{array}$
	\end{center}
where $S=\{q_2', q_2''\}$ and $i=1,2$. The initial state of $\hat{N}$ is $(q_1,\{q_2\},q_2)$.
Obviously, $\hat{N}$ computes the relation $\mathcal{R}(T_1)\circ \mathcal{R} (T_2)$.

In the following, we briefly explain our idea.
In a translation of $\hat{N}$ on input $a(e)$,
the subtree $e$ is processed by
$(q_1,S,q_2')$ and $(q_1,S,q_2'')$. 
Note that in a translation of  $\hat{\tau}$ the states $q_2'$ and $q_2''$ would process the same
tree produced by $(q_1,S)$ on input $e$.
Consider the state $(q_1,S,q_2'')$. If $(q_1,S,q_2'')$, when reading $e$,  makes a 
\emph{valid} guess, i.e., $(q_1,S,q_2'')$ guesses a tree $t$ that is producible by
$(q_1,S)$ on input $e$, then $t\in \text{dom}(q_2')$ by construction of $\hat{T}_1$.
Due to previous considerations (cf. Example~\ref{plausible}),
 it is thus sufficient to ensure that all guesses of states of $\hat{N}$
are valid. While obviously in the case of $\hat{N}$, all guesses are indeed valid,
guesses of transducers obtained from the p-construction are in general not always valid; in particular if \emph{deleting rules} are involved.

To be more specific, 
consider the following transducers $T_1'$ and $T_2'$.
Let $T'_1$ contain the rules
\begin{center}
	$\begin{array}{lclclcl}
	q_1(a(x_1,x_2)) &\rightarrow & b(q'_1 (x_1), q_1''(x_2), q_1'''(x_2)) & \quad &  q'_1(e) &\rightarrow & e\\
	\end{array}$
\end{center}
where  $\text{dom} (q_1'')$ consists of all trees whose left-most leaf is labeled by $e$
while $\text{dom} (q_1''')$ consists of all trees whose left-most leaf is labeled by $c$.
Let $T'_2$ contain the rules
\begin{center}
	$\begin{array}{lclclcl}
	q_2(b(x_1,x_2,x_3)) &\rightarrow & q_2(x_1) & \quad &  q_2(e) &\rightarrow & e_j \mid j=1,2.\\
	\end{array}$
\end{center}
As the translation of $T'_1$ is empty, obviously the translation of $\tau'=T'_1\ci T'_2$ is empty as well.
Thus, $\tau'$ is functional. 
However, the p-construction yields the transducer $N'$ with the rules
\begin{center}
	$\begin{array}{lclclcl}
	(q_1,q_2) (a(x_1,x_2)) &\rightarrow & (q_1',q_2) (x_1) & \quad &  (q'_1,q_2)(e) &\rightarrow & e_j \mid j=1,2\\
	\end{array}$
\end{center}
Even though $\tau'=T_1'\ci T_2'$ is functional, the transducer $N'$
is not.
More precisely, on input $a(e,s)$, where $s$ is an arbitrary tree, $N'$
can produce either $e_1$ or $e_2$ while $\tau'$ would produce nothing.
The reason is that in
the translation of $N'$, the tree $a(e,s)$ is processed by the state $(q_1,q_2)$ by
applying the deleting rule $\eta= (q_1,q_2) (a(x_1,x_2)) \rightarrow  (q_1',q_2) (x_1)$.
Applying $\eta$  means that $(q_1,q_2)$ guesses that on input $a(e,s)$, the state $q_1$
 produces a tree of the form $b(t_1,t_2,t_3)$  by applying the rule
$q_1(a(x_1,x_2)) \rightarrow  b(q'_1 (x_1), q_1''(x_2), q_1'''(x_2))$ of $T_1$. 
However, this guess is not valid, i.e., $q_1$ does not produce such a tree on input  $a(e,s)$,
as by definition
$s\notin \text{dom} (q_1'')$ or $s\notin \text{dom} (q_1''')$.
The issue is that $N'$ itself cannot verify the validity of
this guess because, due to the deleting rule $\eta$, $N'$ does not read $s$.

As the reader might have guessed our idea is that the
validity of each guess is verified using look-ahead. First, we need to define
look-ahead.

A \emph{transducer with look-ahead}
(or \emph{la-transducer}) $M'$ is a transducer that is equipped with an automaton
called the \emph{la-automaton}.
Formally, $M'$ is a tuple
$M'= (Q,\Sigma, \Delta, R, q, B)$ where $Q$, $\Sigma$, $\Delta$ and $q$ are defined as for
transducers and $B$ is the la-automaton.
The rules of $R$ are of the form $q(a(x_1\!:l_1,\dots,x_k\!:l_k)) \to t$
where for $i\in [k]$, $l_i$ is a state of $B$.
Consider the input $s$. The la-transducer $M'$ processes $s$ in two phases:
First each input node of $s$ is annotated by the states of $B$ at its children, i.e., an input node $v$ labeled by  $a\in \Sigma_k$ is relabeled by $\langle a,l_1,\dots , l_k \rangle$
if $B$ arrives in the state $l_i$  when processing the $i$-th subtree of $v$.
Relabeling the nodes $s$ provides $M'$ with additional information about 
the subtrees of $s$, e.g.,
if the node $v$ is relabeled by  $\langle a,l_1,\dots , l_k \rangle$ then the
$i$-th subtree of $v$ is a tree in $\text{dom} (l_i)$. 
The relabeled tree is then processed by $M'$.
To this end a rule $q(a(x_1\!:l_1,\dots,x_k\!:l_k)) \to t$ is interpreted as
$q(\langle a,l_1,\dots , l_k \rangle (x_1,\dots,x_k)) \to t$.

In our example, the idea is to equip $N'$ with an la-automaton to verify the validity of guesses. 
In particular, the la-automaton is the domain automaton $A'$ of $T_1'$.
Recall that a state of $A'$ is a set consisting of states of $T_1'$.
To process relabeled trees the rules of $N'$ are as follows
\begin{center}
	$\begin{array}{lclclcl}
	(q_1,q_2) (a(x_1\!:\{q_1'\} ,x_2\!:\{q_1'', q_1''\} )) &\rightarrow & (q_1',q_2) (x_1) & \quad &  (q'_1,q_2)(e) &\rightarrow & e_j \mid j=1,2\\
	\end{array}$
\end{center}
Consider the tree $a(e,s)$, where $s$ is an arbitrary tree. The idea is that if the root of
$a(e,s)$ is relabeled by $\angl{a,\{q_1'\}, \{q_1'', q_1'''\} }$, then due to Lemma~\ref{Property of A},
$e \in \text{dom} (q_1')$ and $s\in \text{dom}(q_1'') \cap \text{dom}(q_1'')$ and thus on input
$a(e,s)$
a tree of the form $b(t_1,t_2,t_3)$ is producible by $q_1$ using
the rule $q_1(a(x_1,x_2)) \rightarrow  b(q'_1 (x_1), q_1''(x_2), q_1'''(x_2))$.
Quite clearly, the root of $a(e,s)$ is \emph{not} relabeled. 
Thus, the translation of $N'$ equipped with the la-automaton $A'$ is empty as the translation of $\tau'$ is.

%% file: mainPartConstruction.tex
\subsection{Construction  of the LA-Transducer $M$}\label{construction}
Recall that for a 
a composition $\tau$ of two transducers $T_1$ and $T_2$,
 we aim to construct an la-transducer $M$ 
such that $M$ is functional if and only if $\tau$ is functional.

In the following 
we show that combining the ideas presented above yields the la-transducer $M$. 
For $T_1$ and $T_2$, we obtain $M$ by first
completing the following steps.
\begin{enumerate}
	\item Construct the domain automaton $A$ of $T_2$
	\item Construct the  transducer $\hat{T}_1$ from $T_1$ and $A$ using the p-construction 
	\item Construct the  transducer $N$ from $\hat{T}_1$ and $T_2$ using the p-construction
\end{enumerate}
We then obtain $M$ by extending $N$ into a transducer with look-ahead.
Note that  the states of $N$ are written as 
$(q,S,q')$ instead of $((q,S),q')$ for better readability, where $(q,S)$ is a state of $\hat{T}_1$
and $q'$ is a state of $T_2$.
Recall that $(q,S,q')$ means that
$q'$ is supposed to process a tree generated by $(q,S)$.
Furthermore, recall that $S$ is a set of states of $T_2$ and that the idea is that
trees produced by $(q,S)$ are  only supposed to be processed by states in $S$.
Thus, we only consider states $(q,S,q')$  of $N$ where $q'\in S$. 

The transducer $M$ with look-ahead is constructed as follows. 
The set of states of $M$ and the initial state 
of $M$ 
are the states
of $N$ and the initial state of $N$, respectively.
The la-automaton of $M$ is the domain automaton $\hat{A}$ of $\hat{T}_1$.

We now define the rules of $M$.
First, recall that a state of $\hat{A}$ is a set consisting of states of $\hat{T}_1$.
Furthermore, 
recall that for a set of right-hand sides $\Gamma$ and a variable $x$, we denote by $\Gamma \s{x})$ 
the set of all states $q$ such that $q (x)$ occurs in some $\gamma \in \Gamma$.
For a right-hand side $\gamma$, the set $\gamma \s{x}$ is defined analogously.
For all rules 
\[
\eta= (q,S,q') (a(x_1,\dots, x_k))\rightarrow \gamma
\]
of $N$ we proceed as follows:
If  $\eta$
is obtained from the rule $(q,S)(a(x_1,\dots, x_k)) \rightarrow \xi$ of $\hat{T}_1$ 
and subsequently translating $\xi$ by the state $q'$ of $T_2$ then we define
the rule
\[
(q,S,q') (a(x_1\!: l_1,\dots, x_k\!: l_k))\rightarrow \gamma
\]
for $M$
where  for $i \in [k]$, $l_i$ is a state of $\hat{A}$ such that $\xi \s{x_i} \subseteq l_i$.
Recall that relabeling a node $v$, that was previously labeled by $a$, by $\angl{a, l_1,\dots, l_k}$ means that
the $i$-th subtree of $v$ is a tree in $\text{dom} (l_i)$. By Lemma~\ref{Property of A},
$s\in \text{dom} (l_i)$ if and only if $s\in \bigcap_{\hat{q}\in l_i} \text{dom} (\hat{q})$.
Thus, if the node $v$ of a tree $s$ is relabeled by $\angl{a, l_1,\dots, l_k}$ then it means that
$(q,S)$ can process subtree of $s$ rooted at $v$ using the rule
$(q,S)(a(x_1,\dots, x_k)) \rightarrow \xi$.

In the following, we present a detailed example for the construction of $M$
for two transducers $T_1$ and $T_2$.
\begin{example}
Let the transducer $T_1$ contain the rules
\begin{center}
	$\begin{array}{lclclcl}
		q_0 (f(x_1,x_2)) & \rightarrow & f (q_1(x_1), q_2 (x_2))     &\quad & q_0 (f(x_1,x_2)) & \rightarrow &  q_3 (x_2)\\
			q_2 (f(x_1,x_2)) & \rightarrow & f (q_2(x_1), q_1 (x_2)) &\quad & q_1 (f(x_1,x_2)) & \rightarrow & f (q_1(x_1), q_1 (x_2)) \\
		q_2 (f(x_1,x_2)) & \rightarrow & f' (q_2(x_1), q_1 (x_2))    &\quad& 	q_1 (f(x_1,x_2)) & \rightarrow & f' (q_1(x_1), q_1 (x_2))\\
		q_2 (e)  & \rightarrow & e                                   &\quad&   q_1 (e) & \rightarrow &  e \\
			q_3 (d) & \rightarrow & d                                &\quad&  q_1 (d) & \rightarrow &  d
	\end{array}$
\end{center}
and let the initial state of $T_1$ be $q_0$.
Informally, when reading the symbol $f$, the states $q_1$ and $q_2$ nondeterministically
decide whether or not to relabel $f$ by $f'$.
However, the domain of $q_2$ only consists of trees whose leftmost leaf is labeled by $e$.
The state $q_3$ only produces the tree $d$ on input $d$.
Thus, the domain of $T_1$ only consists of trees of the form $f(s_1,s_2)$
where $s_1$ and $s_2$ are trees and either the leftmost leaf of $s_2$ is $e$ or $s_2=d$.

The  initial state of the transducer $T_2$ is $\hat{q}_0$ and $T_2$ contains the rules
\begin{center}
$\begin{array}{lclclcl}
\hat{q}_0 (f (x_1,x_2)) &\rightarrow & f (\hat{q}_1 (x_1), \hat{q}_2 (x_1))  & \quad & \hat{q}_1 (f(x_1,x_2)) &\rightarrow & f (\hat{q}_1 (x_1), \hat{q}_1 (x_2))\\
 \hat{q}_0 (d) &\rightarrow & d & \quad &      \hat{q}_1 (f'(x_1,x_2)) & \rightarrow & f' (\hat{q}_1 (x_1), \hat{q}_2 (x_2))\\
     \hat{q}_2 (f(x_1,x_2)) & \rightarrow & f (\hat{q}_2 (x_1), \hat{q}_2 (x_2))                                    & \quad &       \hat{q}_1 (e) & \rightarrow & e \\
     \hat{q}_2 (e) & \rightarrow & e                                    & \quad &      \hat{q}_1 (d) & \rightarrow & d \\
      \hat{q}_2 (d) & \rightarrow & d .\\
\end{array}$
\end{center}
Informally, on input $s$, the state $\hat{q}_2$ produces $s$ if the symbol $f'$ does not
occur in $s$; otherwise $\hat{q_2}$ produces no output.
The state $\hat{q}_1$ realizes the identity.
Hence, the domain of $T_2$ only consists of the tree $d$ and trees $f(s_1,s_2)$ with no occurrences of $f'$
in $s_1$.  

Consider the composition $\tau=T_1\ci T_2$.
On input $s$, the composition $\tau$ yields $f(s_1,s_1)$ if  $s$ is of the form 
$f(s_1,s_2)$
 and the leftmost leaf of $s_2$ is labeled by
$e$. If the input tree is of the form $f(s_1,d)$, the output tree $d$ is produced.
Clearly, $\tau$ is functional.
We remark that both phenomena described  in Section~\ref{mainpartintro}  occur in the composition $\tau$.
More precisely, simply applying the p-construction to $T_1$ and $T_2$ yields a nondeterministic transducer 
due to  `independent guessing'.
Furthermore, not checking the validity of guesses causes nondeterminism on input $f(s_1,d)$.

In the following, we show how to construct the la-automaton $M$ from the transducers $T_1$ and $T_2$.

\medskip\textbf{Construction of the domain automaton $\mathbf{A}$.}
We begin by constructing the domain automaton
$A$ of $T_2$. The set of states of $A$  is the power set of the set of states of $T_2$ and
the initial state of $A$ is $\{\hat{q}_0\}$.
The rules of $A$ are
\begin{center}
	$\begin{array}{rlclc rlcl}
	\{\hat{q}_0\} & (f (x_1,x_2)) &\rightarrow & f (S (x_1), \emptyset (x_2))  \\
	\{\hat{q}_0\} & (d) &\rightarrow & d\\
	S  &(f(x_1,x_2)) &\rightarrow & f ( S (x_1), S (x_2)) \\
	S  & (e) & \rightarrow & e \\                                        
	S & (d) & \rightarrow & d \\
	\end{array}$
\end{center}
where $S=\{\hat{q}_1,\hat{q_2}\}$. The state $\emptyset$ realizes the identity.
The rules for the state $\emptyset$ are straight forward and hence omitted here. 
 All remaining states, such as for instance $\{\hat{q}_0, \hat{q}_1\}$,
  are unreachable and hence the corresponding rules are irrelevant.
Thus, we omit these rules as well. In the following, we only consider rules of states that are reachable.

\medskip\textbf{Construction of the transducer $\mathbf{\hat{T}_1}$.} For $T_1$ and $A$, the p-construction
yields the transducer $\hat{T}_1$. The transducer $\hat{T}_1$ contains the rules
\begin{center}
	$\begin{array}{rlcl c rlcl}
	(q_0, \{\hat{q}_0\}) & (f(x_1,x_2)) & \rightarrow & f ( (q_1, S )(x_1), q_2 (x_2)) \\
	(q_0, \{\hat{q}_0\}) & (f(x_1,x_2)) & \rightarrow & (q_3,  \{\hat{q}_0\}) (x_2)  \vspace{0.12cm}\\
	
	q_1 & (f(x_1,x_2)) & \rightarrow & f ( q_1(x_1), q_1 (x_2)) \\  
	q_1 &(f(x_1,x_2)) & \rightarrow & f' ( q_1(x_1), q_1 (x_2)) \\ 
	q_1& (e) & \rightarrow &  e \\
	q_1 & (d) & \rightarrow &  d  \vspace{0.12cm}\\
	
  (q_1,S) &(f(x_1,x_2)) & \rightarrow & f ( (q_1,S)(x_1), (q_1, S) (x_2))\\
  (q_1,S)& (e) & \rightarrow &  e \\
  (q_1,S) & (d) & \rightarrow &  d \vspace{0.12cm}\\
  
  	q_2  &(f(x_1,x_2)) & \rightarrow & f (q_2(x_1), q_1 (x_2))                    \\ 
  	q_2 & (f(x_1,x_2)) & \rightarrow & f' (q_2(x_1), q_1 (x_2))    \\    
  	q_2 &(e)  & \rightarrow & e \vspace{0.12cm}\\
  	
  	(q_3,  \{\hat{q}_0\})& (d)  & \rightarrow & d  \\
	\end{array}$
\end{center}
and the initial state of  $\hat{T}_1$ is $(q_0, \{\hat{q}_0\})$.
For better readability, we just write $q_1$ and $q_2$ instead of $(q_1,\emptyset)$ 
and $(q_2,\emptyset)$, respectively.

\medskip\textbf{Construction of the transducer $\mathbf{N}$.} 
For $\hat{T_1}$ and $T_2$,
we construct the transducer $N$  containing the rules
\begin{center}
	$\begin{array}{rlclcrlcl}
	(q_0, \{\hat{q}_0\}, \hat{q}_0) & (f(x_1,x_2)) & \rightarrow & f ( (q_1, S, \hat{q}_1)(x_1), (q_1, S, \hat{q}_2)(x_1)  )\\
	(q_0, \{\hat{q}_0\}, \hat{q}_0) & (f(x_1,x_2)) & \rightarrow & (q_3,  \{\hat{q}_0\}, \hat{q}_0) (x_2)\vspace{0.12cm}\\
	
	(q_1,S,\hat{q}_1) &(f(x_1,x_2)) & \rightarrow & f ( (q_1,S,\hat{q}_1)(x_1), (q_1, S, \hat{q}_1) (x_2)) \\
	  (q_1,S, \hat{q}_1) & (e) & \rightarrow &  e \\
	(q_1,S, \hat{q}_1) & (d) & \rightarrow &  d \vspace{0.12cm}\\
	
	(q_1,S, \hat{q}_2)& (f(x_1,x_2)) & \rightarrow & f ( (q_1,S, \hat{q}_2)(x_1), (q_1, S, \hat{q}_2) (x_2)) \\
	 (q_1,S, \hat{q}_2) & (e) & \rightarrow &  e \\
	(q_1,S, \hat{q}_2) &(d) & \rightarrow &  d \vspace{0.12cm}\\
	
	  	(q_3,  \{\hat{q}_0\}, \{\hat{q}_0\})& (d)  & \rightarrow & d  \\
\end{array}$
\end{center}
The initial state of $N$ is $(q_0, \{\hat{q}_0\}, \hat{q}_0)$.
Note that  the states such as $(q_1,S,\hat{q}_0)$ are not considered as $\hat{q}_0$ is not contained in $S$.
We remark that though no nondeterminism is caused by `independent guessing', $N$ is still nondeterministic on input $f(s_1,d)$
as the validity of guesses cannot be checked.
To perform validity checks for guesses, we extend $N$ with look-ahead. 

\medskip\textbf{Construction of the look-ahead automaton $\mathbf{\hat{A}}$.}
Recall that the look-ahead automaton of $M$ is the domain automaton $\hat{A}$ of $\hat{T}_1$.
The set of states of $\hat{A}$ is the power set of the set of states of $\hat{T_1}$.
The initial state of $\hat{A}$ is $\{ (q_0, \{\hat{q}_0\}) \}$ and $\hat{A}$ contains the following rules.
\begin{center}
	$\begin{array}{rlclcrlcl}
	\{ (q_0, \{\hat{q}_0\}) \} & (f(x_1,x_2)) & \rightarrow & f ( \{(q_1, S )\} (x_1), 
		\{q_2\} ) \} (x_2)) \\
	\{ (q_0, \{\hat{q}_0\}) \} & (f(x_1,x_2)) & \rightarrow &  f ( \emptyset (x_1),  \{(q_3,  \{\hat{q}_0\}) \} (x_2)) \vspace{0.12cm}\\
	
	\{ q_1 \} & (f(x_1,x_2)) & \rightarrow & f ( \{ q_1 \} (x_1), \{ q_1 \} (x_2)) & \quad &  \{q_1\} &(e) & \rightarrow &  e \\
 \{ q_1 \} & (d) & \rightarrow &  d  \vspace{0.12cm}\\
	
	\{ (q_1,S) \} & (f(x_1,x_2)) & \rightarrow & f ( \{ (q_1,S)(x_1) \}, \{ (q_1, S)\} (x_2)) \\
	 \{ (q_1,S) \}& (e) & \rightarrow &  e \\
	\{ (q_1,S) \} & (d) & \rightarrow &  d \vspace{0.12cm}\\
	
	\{ q_2 \} & (f(x_1,x_2)) & \rightarrow & f ( \{ q_2 \}(x_1), \{ q_1 \} (x_2)) \\
	  \{ q_2 \}& (e)  & \rightarrow & e \vspace{0.12cm}\\
	
		(q_3,  \{\hat{q}_0\}, \{\hat{q}_0\})& (d)  & \rightarrow & d  \\
\end{array}$
\end{center}
For better readability, we again just write $q_1$ and $q_2$ instead of $(q_1,\emptyset)$ 
and $(q_2,\emptyset)$, respectively.
We remark that, by construction of the domain automaton, $\hat{A}$ also contains the rule
\[
\{ (q_0, \{\hat{q}_0\}) \}  (f(x_1,x_2)) \rightarrow  f ( \{(q_1, S )\} (x_1), \{q_2, (q_3,  \{\hat{q}_0\}) \} (x_2)) ,
\]
however,  since no rules are defined for the state $\{q_2, (q_3,  \{\hat{q}_0\}) \}$, this rule can be omitted.

\medskip\textbf{Construction of the la-transducer $\mathbf{M}$.} Finally, we construct the la-transducer $M$.
The initial state of $M$ is $(q_0, \{\hat{q}_0\}, \hat{q}_0)$ and the rules of $M$ are\\
\begin{center}
	$\begin{array}{rlclcrlcl}
	(q_0, \{\hat{q}_0\}, \hat{q}_0) & (f(x_1 \!\!:\! \{ (q_1,S) \} ,x_2\!\!: \! \{q_2\})) & \rightarrow & f ( (q_1, S, \hat{q}_1)(x_1), (q_1, S, \hat{q}_2)(x_1)  )\\
	(q_0, \{\hat{q}_0\}, \hat{q}_0) & (f(x_1 \!\!:\! \emptyset ,x_2\!\!: \! \{q_3,  \{\hat{q}_0\} \})) & \rightarrow & (q_3,  \{\hat{q}_0\}, \hat{q}_0) (x_2)\vspace{0.12cm} \\
	
	(q_1,S,\hat{q}_1) &(f(x_1\!\!:\! \{(q_1,S)\},x_2\!\!:\! \{(q_1,S)\})) & \rightarrow & f ( (q_1,S,\hat{q}_1)(x_1), (q_1, S, \hat{q}_1) (x_2))\\
	(q_1,S, \hat{q}_1) & (e) & \rightarrow &  e \\
	(q_1,S, \hat{q}_1) & (d) & \rightarrow &  d \vspace{0.12cm}\\
	
	(q_1,S, \hat{q}_2)& (f(x_1\!\!:\! \{(q_1,S)\},x_2\!\!:\! \{(q_1,S)\})) & \rightarrow & f ( (q_1,S, \hat{q}_2)(x_1), (q_1, S, \hat{q}_2) (x_2))\\
	(q_1,S, \hat{q}_2) & (e) & \rightarrow &  e \\
	(q_1,S, \hat{q}_2) &(d) & \rightarrow &  d  \vspace{0.12cm}\\
	
	 	(q_3,  \{\hat{q}_0\}, \{\hat{q}_0\})& (d)  & \rightarrow & d  \\
\end{array}$
\end{center}
\noindent
By construction, the transducer $N$ contains the rule 
\[\eta=(q_0, \{\hat{q}_0\}, \hat{q}_0)  (f(x_1,x_2))  \rightarrow f ( (q_1, S, \hat{q}_1)(x_1), (q_1, S, \hat{q}_2)(x_1)  ).\]  This rule is obtained from
the rule 	
$(q_0, \{\hat{q}_0\}) (f(x_1,x_2)) \rightarrow f ( (q_1, S )(x_1), q_2 (x_2))$ of $\hat{T}_1$.

Consider the input tree $f(s_1,s_2)$ where $s_1$ and $s_2$ are arbitrary ground trees.
Clearly, translating $f(s_1,s_2)$ with $N$ begins with the rule $\eta$.
Recall that the transducer $N$ is equipped with look-ahead in order
to guarantee that guesses performed by states of $N$ are valid.
In particular, to  guarantee that the guess corresponding to
$\eta$ is valid, we need to test whether or
not $s_1 \in \text{dom} (q_1,S)$ and  $s_2 \in \text{dom} (q_2)$.
Therefore, $M$ contains the rule
\[
(q_0, \{\hat{q}_0\}, \hat{q}_0) (f(x_1 \!\!:\! \{(q_1,S)\} ,x_2\!\!: \! \{q_2\})) \rightarrow  f ( (q_1, S, \hat{q}_1)(x_1), (q_1, S, \hat{q}_2)(x_1)  ).
\]
Recall that if $f$ is relabeled by $\langle f, \{q_1,S\}, \{q_2\} \rangle$ via the la-automaton
$\hat{A}$, this means precisely that $s_1 \in \text{dom} (q_1,S)$ and  $s_2 \in \text{dom} (q_2)$.
We remark that by definition, $M$ also contains rules of the form 
\[
(q_0, \{\hat{q}_0\}, \hat{q}_0) (f(x_1 \!\!:\! l_1 ,x_2\!\!: \! l_2)) \rightarrow  f ( (q_1, S, \hat{q}_1)(x_1), (q_1, S, \hat{q}_2)(x_1)  ), 
\]
where $l_1$ and $l_2$ are states of $\hat{A}$ such that  
$\{q_1,S\} \subseteq l_1$ and $\{q_2\} \subseteq l_2$ and $l_1$ or $l_2$ is a proper superset.
However, as none such states $l_1$ and $l_2$  are reachable by
$\hat{A}$, we have omitted  rules of this form. Other rules are omitted for the same reason.$\hfill\qed$
\end{example}

\subsection{Correctness of the LA-Transducer $M$}
In the following we prove the correctness of our construction. More precisely, we prove that
$M$ is functional if and only if $T_1 \ci T_2$ is. By Lemma~\ref{T1' property 4}, 
it is sufficient to show that $M$ is functional if and only if $\hat{T}_1 \ci T_2$ is.

First, we prove that the following claim:
  If $M$ is functional then $\hat{T}_1 \ci T_2$ is functional.
More precisely, we show that $\mathcal{R} (\hat{T}_1) \circ \mathcal{R} (T_2) \subseteq \mathcal{R} (M)$. Obviously, this
implies our claim. 
First of all, consider the transducers $N$ and $N'$ obtained from the p-construction in our examples in Section~\ref{mainpartintro}.
Notice that the relations defined by  $N$ and $N'$ are supersets of 
$\mathcal{R} (T_1) \circ \mathcal{R} (T_2)$ and $\mathcal{R} (T_1')\circ \mathcal{R}(T_2')$, respectively. 

In the following, we show that this observation can be generalized.
Consider arbitrary transducers $T$ and $T'$.
We claim that the transducer $\breve{N}$ obtained from the p-construction 
for $T$ and $T'$
always defines a superset of the composition $\mathcal{R}(T)\circ \mathcal{R}(T')$.
To see that our claim holds, consider a translation of $T\ci T'$ in which the state $q'$ of $T'$ processes a tree $t$
produced by the state $q$ of $T$ on input $s$. If the corresponding state $(q,q')$ of
$\breve{N}$ processes $s$ then $(q,q')$ can guess that $q$ has produced $t$ and
proceed accordingly.
Thus $\breve{N}$ can effectively simulate the composition  $T\ci T'$.

As $M$ is in essence obtained from the p-construction extended with look-ahead,
$M$ `inherits' this property.
Note that the addition of look-ahead does not affect this property.
Therefore our claim follows.

\begin{lemma}\label{subset}
	$\mathcal{R}(\hat{T}_1) \circ \mathcal{R}(T_2) \subseteq \mathcal{R}(M)$.
\end{lemma}

\noindent
In fact an even stronger result holds.

\begin{lemma}\label{stronger subset claim}
	Let $(q_1,S)$ be a state of $\hat{T}_1$ and $q_2$ be a state of $T_2$.
	If on input $s$, $(q_1,S)$ can produce the tree $t$ and on input $t$, $q_2$ can produce the tree $r$
	then $(q_1,S,q_2)$ can produce $r$ on input $s$.
\end{lemma}

Consider a translation of $\hat{T}_1\ci T_2$ in which  $T_2$ processes the tree $t$
produced by $T_1$ on input $s$.
We call a translation of $M$ \emph{synchronized} if the translation simulates a translation of
$\hat{T}_1 \ci T_2$, i.e.,
if a state $(q,S,q')$ of $M$ processes the subtree $s'$ of $s$
and the corresponding state of $q'$ of $T_2$ processes the subtree $t'$ of $t$ and $t'$ is produced by
$(q,S)$ on input $s'$,
then $(q,S,q')$ guesses $t'$.

We now show that if $\hat{T}_1 \ci T_2$ is functional, then so is $M$.
Before we prove our claim consider the following auxiliary results.

\begin{lemma}\label{aux}
	Consider an arbitrary input tree $s$. Let $\hat{s}$ be a subtree of $s$.
	Assume that in an arbitrary translation of $M$ on input $s$, 
	the state $(q_1,S,q_2)$ processes $\hat{s}$. Then, a synchronized translation of $M$ on input $s$
	exists in which the  state $(q_1,S,q_2)$ processes the subtree $\hat{s}$.
\end{lemma}

\noindent
It is easy to see that the following result holds for arbitrary transducers.
\begin{proposition}\label{prop}
	Let $\tau=T_1 \ci T_2$ where $T_1$ and $T_2$ are arbitrary transducers.
	Let $s$ be a tree such that $\tau (s)=\{r\}$ is a singleton.
	Let $t_1$ and $t_2$ be distinct trees produced by $T_1$ on input $s$.
	If $t_1$ and $t_2$ are in the domain of $T_2$ then
	$T_2(t_1) =T_2 (t_2) =\{r\}$.
\end{proposition}
\noindent
Using Lemma~\ref{aux} and Proposition~\ref{prop}, we now show that the following holds.
Note that in the following $t/v$, where $t$ is some tree and $v$ is a node, denotes the
subtree of $t$ rooted at the node $v$.

\begin{lemma}\label{main lemma}
	Consider an arbitrary input tree $s$.  Let $\hat{s}$ be a subtree of $s$.
	Let the state $(q_1,S,q_2)$ process $\hat{s}$
	in a translation $M$ on input $s$.
	If $\hat{T}_1\ci T_2$ is functional then $(q_1,S,q_2)$ can only produce
	a single output tree on input $\hat{s}$.
\end{lemma}
\begin{proof*}
	Assume to the contrary that $(q_1,S,q_2)$ can produce distinct trees $r_1$ and $r_2$ on input $\hat{s}$.
	For $r_1$, it can be shown that a tree $t_1$ exists such that
	\begin{enumerate}
		\item on input $\hat{s}$, the state $(q_1,S)$ of $\hat{T}_1$ produces $t_1$ and
		\item on input $t_1$, the state $q_2$ of $T_2$ produces $r_1$.
	\end{enumerate}
It can be shown that a tree $t_2$ with the same properties exists for $r_2$.
Informally, this means that
$r_1$ and $r_2$ are producible by $(q_1,S,q_2)$ by simulating the `composition of $(q_1,S)$ 
and $q_2$'.

Due to Lemma~\ref{aux}, a  synchronized translation of $M$ on input $s$
exists in which the  state $(q_1,S,q_2)$ processes the subtree $\hat{s}$ of $s$.
Let $g$ be the node at which $(q_1,S,q_2)$ processes $\hat{s}$.
Let $\hat{q}_1,\dots, \hat{q}_n$ be all states of $M$ of the form $(q_1,S,q_2')$,
where $q_2'$ is some state of $T_2$, that occur in the synchronized translation of
$M$ and that process  $\hat{s}$.
Note that by definition $q_2' \in S$.
Due to Lemmas~\ref{Property of hatT1} and~\ref{stronger subset claim}, 
we can assume that in the synchronized translation, the states $\hat{q}_1,\dots, \hat{q}_n$ all guess that the tree
$t_1$ has been produced by the state $(q_1,S)$ of $\hat{T}_1$ on input $\hat{s}$.
Hence, we can assume that at the node $g$, the output subtree $r_1$ is produced.
Therefore, a synchronized translation of $M$ on input $s$ exists, that yields an output tree $\hat{r}_1$ such that
$\hat{r}_1 /g= r_1$, where $\hat{r}_1 /g$ denotes the subtree of $\hat{r}_1$ rooted at the node $g$.
Analogously, it follows that a synchronized translation of $M$ on input $s$ exists, that yields an output tree $\hat{r}_2$ such that
$\hat{r}_2 /g= r_2$. 

As both translation are synchronized, i.e., `simulations' of translations of $\hat{T}_1\ci T_2$ on input $s$, it follows that
the trees  $\hat{r}_1$ and  $\hat{r}_2$ are producible by $\hat{T}_1\ci T_2$ on input $s$. Due to Proposition~\ref{prop},
$\hat{r}_1=\hat{r}_2$ and therefore $r_1=\hat{r}_1/g=\hat{r}_2/g=r_2$.
\end{proof*}

Lemma~\ref{subset} implies that if $M$ is functional then $\hat{T}_1\ci T_2$ is functional as well.
Lemma~\ref{main lemma} implies that if $\hat{T}_1\ci T_2$ is functional then so is $M$.
Therefore, we deduce that due Lemmas~\ref{subset} and~\ref{main lemma} the following holds.

\begin{corollary}\label{main}
	$\hat{T}_1\ci T_2$ is functional if and only if $M$ is functional.
\end{corollary}
In fact, Corollary~\ref{main} together with Lemma~\ref{subset}
imply that $\hat{T}_1\ci T_2$ and $M$ are equivalent
 if $\hat{T}_1\ci T_2$ is functional, since
it can be shown that $\text{dom}(\hat{T}_1\ci T_2) = \text{dom} (M)$.

Since functionality for transducers with look-ahead is decidable~\cite{DBLP:journals/actaC/Esik81},
Corollary~\ref{main} implies that it is decidable whether or not  $\hat{T}_1\ci T_2$ is functional.
Together with Lemma~\ref{T1' property 4}, we obtain:
\begin{theorem}\label{main theorem 1}
	Let $T_1$ and $T_2$ be top-down tree transducers.
	It is decidable whether or not $T_1\ci T_2$ is functional.
\end{theorem}

%% file: CompositionNTransducers.tex
\subsection{Functionality of Arbitrary Compositions}
In this section, we show that the question whether or not an arbitrary 
composition is functional can be reduced to the question of
whether or not a two-fold composition is functional.

\begin{lemma}\label{composition of n transducers lemma 2}
Let $\tau$ be a composition of transducers.
Then two transducers $T_1,T_2$ can be constructed such that
$T_1\ci T_2$ is functional if and only if $\tau$ is functional.
\end{lemma}

\begin{proof*}
	Consider a composition of $n$ transducers  $T'_1,\dots, T'_n$.
	W.l.og. assume that $n>2$. For $n\leq 2$, our claim follows trivially.
	Let $\tau$ be the composition of  $T'_1,\dots, T'_n$.
	We show that transducer $\hat{T}_1,\dots, \hat{T}_{n-1}$ exist such that
	$\hat{T}_1 \ci \cdots \ci \hat{T}_{n-1}$ is functional if and only if $\tau$ is.
	
	Consider an arbitrary input tree $s$. Let $t$ be a tree produced by the composition
	$T'_1 \ci \cdots \ci T'_{n-2}$ on input $s$.
	Analogously as in Proposition~\ref{prop}, the composition $T'_{n-1}\ci  T'_n$, on input $t$, can only produce a single output tree
	if $\tau$ is  functional.
	For the transducers $T'_{n-1}$ and  $T'_{n}$, we construct the la-transducer $M$
	according to our construction in Section~\ref{construction}.
	It can be shown that, the la-transducer $M$  our construction yields has the following properties
	regardless of whether or not $T_{n-1}\ci T_{n}$ is functional
	\begin{enumerate}
		\item[(a)] $\text{dom}(M)=\text{dom}(T_{n-1}\ci T_{n})$  and
		\item[(b)] on input $t$, $M$ only produces a single output tree if and only if $T_{n-1}\ci T_{n}$ does
	\end{enumerate}
	Therefore,
	$\tau(s)$ is a singleton if and only if 
	$T'_1 \ci \cdots \ci T'_{n-2}\ci M (s)$ is a singleton.
	Engelfriet has shown that every transducer with look-ahead can be decomposed to a composition of a 
	deterministic bottom-up relabeling and a transducer~(Theorem 2.6 of~\cite{DBLP:journals/mst/Engelfriet77}).
	It is well known that (nondeterministic) relabelings are
	independent of whether they are defined by bottom-up transducers or by top-down transducers~(Lemma~ 3.2 of \cite{DBLP:journals/mst/Engelfriet75}).
	Thus, any  transducer with look-ahead can be decomposed into a composition of a 
	nondeterministic top-down relabeling and a transducer.
	Let $R$ and $T$ be the relabeling and the transducer such that $M$ and $R\ci T$
	are equivalent. 
	Then obviously,  $\tau (s)$ is a singleton if and only if 
	$T'_1 \ci \cdots \ci T'_{n-2}\ci R\ci T (s)$ is a singleton.
	
	Consider arbitrary transducers $\bar{T}_1$ and $\bar{T}_2$.
	Baker has shown that  if
	$\bar{T}_2$ is non-deleting and linear then
	a transducer $T$ can be constructed such that $T$ and $\bar{T}_1\ci \bar{T}_2$
	are equivalent~(Theorem 1 of~\cite{DBLP:journals/iandc/Baker79b}).
	By definition, any relabeling is non-deleting and linear.
	Thus, we can construct a transducer $\tilde{T}$ such that $\tilde{T}$ and $T'_{n-2}\ci R$
	are equivalent.
	Therefore, it follows that  $\tau(s)$ is a singleton if and only if 
	$T'_1 \ci \cdots \ci T'_{n-3}\ci \tilde{T}\ci T (s)$ is a singleton.
	This yields our claim.
\end{proof*}

\noindent
Lemma~\ref{composition of n transducers lemma 2} and Theorem~\ref{main theorem 1} yield that
functionality of compositions of transducers is decidable.

Engelfriet has shown that any la-transducer can be decomposed into a composition of a
nondeterministic top-down relabeling and a transducer~\cite{DBLP:journals/mst/Engelfriet77,DBLP:journals/mst/Engelfriet75}.
Recall that while la-transducers  generalize transducers,
bottom-up transducers and la-transducers are incomparable~\cite{DBLP:journals/mst/Engelfriet77}.
Baker, however, has shown that the composition of $n$ bottom-up-transducers can be realized by
the composition of $n+1$ top-down transducers~\cite{DBLP:journals/iandc/Baker79b}.
For any functional composition of transducers an equivalent  deterministic la-transducer can be constructed~\cite{DBLP:journals/ipl/Engelfriet78}. 
Therefore we obtain our following main result.

\begin{theorem}\label{main 2}
Functionality for arbitrary
compositions of top-down and bottom-up tree transducers 
is decidable.
In the affirmative case, an equivalent deterministic top-down tree transducer
with look-ahead can be constructed. 
\end{theorem}

%% file: conclusion.tex
\section{Conclusion}
We have presented a construction of an la-transducer for a composition of transducers which is functional if and only if
the composition of the transducers is functional --- in which case it is equivalent to the composition.
This construction is remarkable since transducers are not closed under composition in general,
neither does functionality of the composition imply that each transducer occurring therein, is functional.
%
By Engelfriet's construction in~\cite{DBLP:journals/ipl/Engelfriet78}, 
our construction provides the key step to an efficient implementation (i.e., a deterministic transducer,
possibly with look-ahead) for a composition of transducers -- whenever possible (i.e., when their translation is functional).
As an open question, it remains to see how large the resulting functional transducer necessarily must be,
and whether the construction can be simplified if for instance only compositions of linear transducers are considered.

%% file: appendix.tex
\section{Appendix}
In the following, we first introduce additional notation and definitions used in
the proofs in the Appendix.

\subsection{Definitions}
\subsubsection{Set of Nodes} Let $t$ be a tree.
For $t$, its set  $V(t)$ of  nodes is a subset of $V=\mathbb{N}^*$.
More formally,
$V(t)=\{\epsilon \} \cup \{iu \mid i\in [k], u\in V(t_i) \}$
where $t=a(t_1,\dots, t_k)$, $a\in \Sigma_k$, $k\geq 0$ and $t_1,\dots,t_k \in T_\Sigma$.
For better readability we add dots between numbers.
E.g. for the tree $t=f(a,f(a,b))$ we have $V(t)=\{\epsilon,1,2,2.1,2.2\}$.
For  $v\in V(t)$, $t[v]$ is the label of $v$ and
$t/v$ is the subtree of $t$ rooted at $v$. 

\subsubsection{Substitutions}
Let $t_1,\dots t_n$ be trees over $\Sigma$ and $v_1,\dots, v_n$ be distinct nodes none of which is a prefix of the other, 
then we denote by $[v_i \leftarrow t_i\mid i\in[n]]$ 
the \emph{substitution} 
that for each $i\in [n]$, replaces the subtree rooted at $v_i$ with $t_i$.

Let $t$ be a tree, $a\in \Sigma_0$ and $\mathcal{T}$ be a set of trees.
We denote by $t[a \leftarrow \mathcal{T}]$ the 
set of all trees obtained by substituting
leaves labeled by $a$  with some tree in $\mathcal{T}$, i.e.,
the set of all trees of the form $t[v\leftarrow t_v \mid v\in V(t), t[v]=a]$ where for all $a$-leaves $v$, $t_v\in \mathcal{T}$.
Note that two distinct leaves labeled by $a$ may be replaced by distinct trees in $\mathcal{T}$.
If $\mathcal{T}=\emptyset$ then we define  $t[a \leftarrow \mathcal{T}]=\emptyset$.
For simplicity, we write $t[a \leftarrow t']$ if $\mathcal{T}=\{t'\}$.

\subsubsection{Partial Trees and Semantic of a Transducer}
Recall that
$T_\Sigma[B]=T_{\Sigma'}$ where $\Sigma'$ is obtained from $\Sigma$
by $\Sigma'_0=\Sigma_0\cup B$ while for all $k>0$, $\Sigma'_k =
\Sigma_k$. In the following, we call a tree in $T_\Sigma[B]$  a \emph{partial tree}.
 
The \emph{semantic} of a transducer $T$, defined as in Section~\ref{top-down}, is formally
defined as follows.
Let $q\in Q$ and $v$ be an arbitrary node.
We denote by $\p{q}^T_v$ the partial function from $T_\Sigma [B]$ to the power set of $T_\Delta [Q (V)]$
defined as follows
\begin{itemize}
	\item for $s=a(s_1,\dots,s_k)$, $a\in \Sigma_k$, and $s_1,\dots,s_k\in T_\Sigma[B]$, 
	\[\p{q}^T_v(s)= \bigcup_{\xi \in\text{rhs}_T(q,a)} \xi[q(x_i) \leftarrow \p{q}_{v.i} (s_i) \mid q\in Q, i\in [k]]\]
	\item for $b\in B$, $\p{q}^T_v(b)= \{ q(v) \}$,
\end{itemize}
where $\text{rhs}_T(q,a)$ denotes the set of all right-hand sides of $q$ and $a$.
The reason why input nodes of $s$ are added to the semantic of $T$ is that for some of our proofs
we require that for states $q$ of $T$ it is traceable which input node $q$
currently processes.

If clear from context which transducer is meant, we omit the superscript $T$ and 
write $\p{q}_v$ instead of $\p{q}^T_v$.
In the following, we write 
$\p{q}$ instead of $\p{q}_\epsilon$ for simplicity. 
We write $\p{q}^T_v(s) \Rightarrow t$ if $t\in \p{q}^T_v(s)$.

In the following, for trees in $T_\Delta [Q(X)]$ and $T_\Delta [Q(V)]$, we write
$t\langle x\leftarrow v\rangle$ to denote the substitution
$t[q(x)\leftarrow q(v)\mid q\in Q]$ for better readability where $x\in X$ and $v\in V$.

Recall that for a set $\Gamma$ of right-hand sides of a transducer $T$,
$\Gamma \s{x_i}$ denotes the set of all states $q$ of $T$ such that
$q(x_i)$ occurs in some tree $\gamma$ in $\Gamma$.
For a set $\Lambda$ of trees  in $T_\Delta [Q (V)]$, we define
$\Lambda \s{v}$ where $v$ is some node analogously;
e.g.,  for
$\Lambda=\{f(q(v_1),f (q(v_2), q'(v_2))),\ f(q_1(v_1), q_2 (v_2))\}$, we have
$\Lambda \s{v_1} =\{q,q_1\}$ and $\Lambda \s{x_2} =\{q,q',q_2\}$.

%% file: appendix-ii.tex
\section{Properties of the Domain Automaton $A$}
In the following we consider the \emph{domain automaton} $A$  introduced in Section~\ref{mainpartintro} for a transducer $T$.
In particular, we consider the properties of $A$. Recall that a state of $A$
is a set consisting of states of $T$.
In  Section~\ref{mainpartintro}, we have claimed that
if in a translation of $T$ on input $s$, 
the states $q_1\dots,q_n$ process the node $v$ of $s$
then $\{q_1\dots,q_n\}$ processes the node $v$ of $s$ in a computation of $A$.
We now formally prove this statement.
First we prove the following auxiliary result.

\begin{lemma}\label{A property 2}
	Let $s\in T_\Sigma [X]$.
	Let $v_1,\dots v_n$ be the nodes of $s$ that are labeled by a symbol in $X$.
	Let $S_1$ and $S_2$ be states of $A$ and let for $j=1,2$,
	\[ \p{S_j} (s) \Rightarrow s[v_i\leftarrow S^i_j (v_i) \mid i\in [n]] \]
	where for $i\in [n]$, $S_j^i$ is a state of $A$.
	Then 	\[ \p{S_1\cup S_2} (s) \Rightarrow s[v_i\leftarrow S^i_1 \cup S^i_2 (v_i) \mid i\in [n]].\]
\end{lemma}

\begin{proof*}
	We prove our claim by structural induction over $s$.
	Let $s=a(s_1,\dots, s_k)$ where $a\in \Sigma_k$, $k\geq 0$, and for $\iota\in [k]$, $s_\iota \in T_\Sigma [X]$.
	As 
    \[ \p{S_1} (s) \Rightarrow s[v_i\leftarrow S^i_1 (v_i) \mid i\in [n]], \]
	a rule
	$S_1 (a (x_1,\dots,x_k)) \rightarrow a(\hat{S}_{1} (x_1),\dots,\hat{S}_{k} (x_k))$ exists such that for all $\iota\in [k]$,
	on input $s_\iota$,
	the function $\p{\hat{S}_\iota}_\iota$ yields
	the subtree of $s[v_i\leftarrow S^i_1 (v_i) \mid i\in [n]]$ that is rooted at the node $\iota$.
	More formally,
	\[ 
	\p{\hat{S}_\iota}_\iota (s_\iota)\Rightarrow 
	s_\iota[v'\leftarrow S^i_1 (\iota.v') \mid v'\in V(s_\iota) \text{ and } \iota.v'=v_i  \text{ where } i\in[n]],
	\] 
	which in turn implies
		\[ 
	\p{\hat{S}_\iota} (s_\iota)\Rightarrow 
	s_\iota[v'\leftarrow S^i_1 (v') \mid v'\in V(s_\iota) \text{ and } \iota.v'=v_i  \text{ where } i\in[n]].
	\] 
	Analogously, it follows that a rule $S_2 (a (x_1,\dots,x_k)) \rightarrow a(\hat{S}'_{1} (x_1),\dots,\hat{S}_{k}' (x_k))$ exists such that for all $\iota\in [k]$,
		\[ \p{\hat{S}'_\iota} (s_\iota)\Rightarrow 
	s_\iota[v'\leftarrow S^i_2 (v') \mid v'\in V(s_\iota) \text{ and } \iota.v'=v_i  \text{ where } i\in[n] ].
	\]
	We now show that  the automaton $A$ contains the rule
	\begin{equation}\tag{a}
	S_1\cup S_2 (a(x_1,\dots, x_k))\rightarrow a(\hat{S}_1\cup \hat{S}'_1 (x_1),\dots , \hat{S}_{k} \cup \hat{S}_{k}' (x_k)).
	\end{equation}
	By construction, the rule 
	$S_1 (a (x_1,\dots,x_k)) \rightarrow a(\hat{S}_{1} (x_1),\dots,\hat{S}_{k} (x_k))$
	is defined only if for all $q\in S_1$ a non-empty set of right-hand sides $\Gamma_q \subseteq
	\text{rhs}_T (q,a)$ exists such that
	for $\iota\in [k]$, $\hat{S}_\iota=\bigcup_{q\in S_1} \Gamma_q \s{x_\iota}$.\\
	Likewise, the rule $S_2 (a (x_1,\dots,x_k)) \rightarrow a(\hat{S}'_{1} (x_1),\dots,\hat{S}_{k}' (x_k))$
	is defined only if for all $q'\in S_2$ a non-empty set of right-hand sides $\Gamma'_{q'} \subseteq
	\text{rhs}_T (q',a)$ exists such that
	for $\iota\in [k]$, $\hat{S}'_\iota=\bigcup_{q'\in S_2} \Gamma'_{q'} \s{x_\iota}$.
	For all states $q\in S_1\cup S_2$, we define
	\[
	\breve{\Gamma}_q = \begin{cases}
	\Gamma_q \cup \Gamma'_q & \text{ if } q\in S_1 \cap S_2 \\
	\Gamma_q & \text{ if } q\in S_1 \setminus S_2 \\
	\Gamma'_q & \text{ if } q\in S_2 \setminus S_1
	\end{cases}
	\]
	Clearly, the sets $\breve{\Gamma}_q$ yield that the rule defined in~(a) exists. 
	
	Now, consider the following.
	As for  $\iota\in [k]$, 	
	\[ \p{\hat{S}_\iota} (s_\iota)\Rightarrow 
	s_\iota[v'\leftarrow S^i_1 (v') \mid v'\in V(s_\iota) \text{ and } \iota.v'=v_i  \text{ where } i\in[n] ]
	\]
	and
	\[ \p{\hat{S}'_\iota} (s_\iota)\Rightarrow 
	s_\iota[v'\leftarrow S^i_2 (v') \mid v'\in V(s_\iota) \text{ and } \iota.v'=v_i  \text{ where } i\in[n] ]
	\]
	the induction hypothesis yields 
	\begin{equation}\tag{b}
	\p{\hat{S}_\iota \cup \hat{S}_\iota'} (s_\iota)\Rightarrow 
	s_\iota[v'\leftarrow S^i_1\cup  S^i_2 (v') \mid v'\in V(s_\iota) \text{ and } \iota.v'=v_i  \text{ where } i\in[n]]
	\end{equation}
	Thus, due to (a) and (b) our claim follows.
\end{proof*}

\noindent
We now prove our statement.

\begin{lemma}\label{A property 3}
	Let $s\in T_\Sigma [X]$. 
	Let $v_1,\dots v_n$ be the nodes of $s$ that are labeled by a symbol in $X$.
	Let $S$ be a state of $A$. For $q\in S$, let
	$\p{q}^T (s) \Rightarrow t_{q}$. 
	Then, 
	 \[\p{S}^A (s) \Rightarrow  s[v_i\leftarrow  S_i(v_i)\mid i\in [n]]\]
	 where $S_i= \bigcup_{q\in S} t_{q}\s{v_i}$.
\end{lemma}

\begin{proof*}
	Due to Lemma~\ref{A property 2}, it is sufficient to show that if
		$\p{q}^T (s) \Rightarrow t_{q}$,
	then
	\[\p{ \{q\}}^A (s) \Rightarrow  s[v_i\leftarrow  S'_i(v_i)\mid i\in [n]]\]
	where $S'_i= t_{q}\s{v_i}$. 
	We prove this claim by structural induction over $s$.
	Let $s=a(s_1,\dots, s_k)$ where $a\in \Sigma_k$, $k\geq 0$, and for $i\in [k]$, $s_i \in T_\Sigma [X]$.
	Due to our premise, $\gamma\in \text{rhs} (q,a)$ exists such that
	\begin{equation}\tag{1}\label{A equation}
	t_q\in \gamma [q'(x_i)\leftarrow \p{q'}_{i} (s_i) \mid q'\in Q, i\in [k]].
	\end{equation}
	By definition of $A$, $\gamma\in \text{rhs} (q,a)$ implies that  the automaton $A$ contains  the rule
	\begin{equation}\tag{a}
	\{q\} (a) \rightarrow a(\hat{S}_1 (x_1),\dots, \hat{S}_k (x_k) )
	\end{equation}
	where $\hat{S}_i=\gamma \s{x_i}$  for  $i\in [k]$.
	
	Now consider $\gamma\in \text{rhs} (q,a)$ in conjunction with the variable $x_i$. Let $\gamma \s{x_i} =\{q_1,\dots, q_m\}$.
	For $j\in [m]$, we denote by $U_j$ the set of all nodes of $\gamma$ that are labeled by
	$q_j (x_i)$.
	For all nodes  $u \in U_j$, 
	Equation~\ref{A equation} clearly implies $\p{q_j}_i^T (s_i) \Rightarrow t_q/u$.
	Recall that by definition, if $\breve{q} (\breve{v})$ occurs in $t_q/u$, where
	$\breve{q}$ is a state of $T$ and $\breve{v}$ is some node then
	$\breve{v}$ is of the form $i.v'$.
	Clearly, $\p{q_j}_i^T (s_i) \Rightarrow t_q/u$ implies 
	$\p{q_j}^T (s_i) \Rightarrow \eta_{u}$ where $\eta_{u}$ denotes the tree
	obtained from $t_q/u$ by substituting occurrences of $\breve{q} (i.v')$
	by $\breve{q} (v')$.

	Recall that $\eta_{u} \s{v'}$ denotes the set of all states $q'$ in $Q$ such that $q'(v')$ occurs in $\eta_{u}$. In the following, let $S_{u,v'}=\eta_{u} \s{v'}$.
	Due to the induction hypothesis, it follows that for all $u\in U_j$,
	\begin{equation}\tag{2}\label{Equation 2}
		\p{ \{q_j\} }^A (s_i) \Rightarrow  s_i[v' \leftarrow S_{u,v'} (v')\mid v'\in V(s_i), s_i[v'] \in X ]. 
	\end{equation}
	Due to Lemma~\ref{A property 2} and Equation~\ref{Equation 2}, it follows for  all $j\in [m]$ that
	\begin{equation}\tag{3}\label{Equation 3}
     \p{ \{q_j\} }^A (s_i) \Rightarrow  s_i[v' \leftarrow \bigcup_{u\in U_j} S_{u,v'}  (v')\mid v'\in V(s_i), s_i[v'] \in X ]
	\end{equation}
	Recall that $\hat{S}_i=\gamma \s{x_i}$.
	Thus, Lemma~\ref{A property 2} and Equation~\ref{Equation 3} yield
	\begin{equation}\tag{b}
	\p{\hat{S}_i}^A (s_i)\Rightarrow  s_i[v' \leftarrow  \bigcup_{j\in [m]} \bigcup_{u\in U_j}  S_{u,v'} (v')\mid v'\in V(s_i), s_i[v'] \in X ]
	\end{equation}
	Note that for $v=i.v'$, 
	\[
	\bigcup_{j\in [m]} \bigcup_{u\in U_j} \eta_{u} \s{v'} = \bigcup_{j\in [m]} \bigcup_{u\in U_j} S_{u,v'}  = t\s{v}.
	\]
	Therefore it follows that (a) and (b) yield our claim.
\end{proof*}

Additionally, the domain automaton $A$ has the following property.
If in a translation of $A$ on input $s$, the state $S$
processes the node $v$ of $s$ then a translation of $T$ on input $s$
exists such that $v$ is only processed by states in $S$.
More formally, we prove thee following result.

\begin{lemma}\label{A property 4}
	Let $s\in T_\Sigma [X]$. 
	Let $v_1,\dots v_n$ be the nodes of $s$ that are labeled by a symbol in $X$.
	Let $S, S_1,\dots, S_n$ be states of $A$. Let $\p{S}^A (s) \Rightarrow \hat{s}$ where
	\[ \hat{s}=  s [v_i \leftarrow S_i (v_i) \mid i\in [n]]. \]
	For all $q \in S$, a tree $t$ exists such that $\p{q}^T (s) \Rightarrow t$
	and  for $i\in [n]$, $t\s{v_i}\subseteq S_i$.
\end{lemma}

\begin{proof*}
	We prove our claim by structural induction.
	Obviously, our claim holds if $s\in X$. 
	
	In the following, let $s\notin X$. 
	Then, a node $v$ exist such that the subtree of $s$ rooted at $v$ is of the form $a(s_1,\dots,s_k)$ where $a\in \Sigma_k$, $k\geq 0$, and $s_1,\dots,s_k \in X$.
	Note that by definition $v$ can be a leaf.	
	Consider the tree $s'=s[v\leftarrow x_1]$. Let
	$v_1',\dots, v_m'$ be the nodes of $s'$ that are labeled by a symbol in $X$.
	W.l.o.g. let $v_1'=v$.
	
	Recall that the node $v$ is labeled by the node $a$ in $s$.
	As $\p{S}^A (s) \Rightarrow \hat{s}$, it follows that
	a tree $\tilde{s}$ exists such that $\p{S}^A (s') \Rightarrow \tilde{s}$
	and the tree $s$ can be obtained from  $\tilde{s}$ by substituting
	$S_1' (v)$ by 	$\xi\langle x_i\leftarrow v.j \mid j\in [k]\rangle$
	where  $\xi$ is a right-hand side of $S_1'$ and $a$.
	Note that by definition, states $S_1',\dots, S_m'$ of $A$ exists such that
	the tree $\tilde{s}$ is obtained from $s'$ by relabeling the node
	$v_i'$ of $s'$ by $S_i' (v_i)$.
	More formally, it holds that
	\[\tilde{s} = s'[v_i'\leftarrow S'_{i} (v_i') \mid i\in [m]], \]
	and that
	\begin{equation}\tag{1}\label{A eqaution}
	 \hat{s}=	\tilde{s}[S_1' (v) \leftarrow
	\xi\langle x_i\leftarrow v.j \mid j\in [k]\rangle 
	].
	\end{equation}
	By induction hypothesis, as $\p{S}^A (s') \Rightarrow \tilde{s}$, for all states $q\in S$, a tree
	$t'$ exists such that 
	\begin{enumerate}
		\item  $\p{q}^T(s') \Rightarrow t'$ and
		\item  for $i\in [m]$, $t'\s{v_i'} \subseteq S'_{i}$.
	\end{enumerate}
	In particular, it holds that $t' \s{v'_1}=t' \s{v} \subseteq S'_{1}$.
	
	Let $S'_{1}=\{q_1,\dots, q_n\}$. 
	Recall that $\xi$ is a right-hand side of $S_1'$ and $a$. 
	W.l.o.g. let $\xi= a(S_1 (x_1), \dots, S_k (x_k))$.
	Then, by definition of the rules of $A$, it follows that for each $j\in [n]$,
	a tree $\gamma_j$ exists such that
	\begin{enumerate}
		\item[(a)]  $\gamma_j \in \text{rhs}_{T} (q_j,a)$ and
		\item[(b)]  for $\iota \in [k]$ it holds that $\bigcup_{j\in [n]} \gamma_j \s{x_\iota} \subseteq S_\iota$. 
	\end{enumerate}
	As $t' \s{v} \subseteq S'_{1}$, it follows that $\p{q}^T (s)\Rightarrow t$
	where
	\begin{equation}\tag{2}\label{A eqaution2}
	t= t'[ q_j (v)  \leftarrow \gamma_j \langle x_i\leftarrow v.i \mid i\in [k]\rangle \mid j\in [n] ].
	\end{equation}
	Consider the node $v.\iota$ where $\iota\in [k]$.
	If $\breve{S} (v.\iota)$ occurs in $\hat{s}$
	then it follows due to Equation~\ref{A equation} that $\breve{S} (x_\iota)$ occurs in $\xi$.
	Due to Equation~\ref{A eqaution2} and Statement~(b), it follows that  
	$t\s{v.\iota} \subseteq S_{\iota}$.  Thus, our claim follows.
\end{proof*}

\noindent
Lemmas~\ref{A property 3} and~\ref{A property 4} yield Lemma~\ref{Property of A}.

%% file: appendix-iii.tex
\section{Properties of $\hat{T}_1$}
In the following we consider the properties of the transducer $\hat{T}_1$.
Recall that $\hat{T}_1$ is obtained via the p-construction from the transducer
$T_1$ and the domain automaton $A$ of $T_2$.
In particular, we formally prove the statements we made about $\hat{T}_1$ in
Section~\ref{mainpartintro}.
First we formally prove Lemma~\ref{Property of hatT1}, that is, we prove the following.

\begin{lemma}\label{T1' property 1}
	Let $(q,S)$ be a state of $\hat{T}_1$ and $S\neq \emptyset$.
	If the tree $t$ over $\Delta$ is producible by $(q,S)$ then
	$t\in_{q_2\in S} \bigcap\text{dom} (q_2)$.
\end{lemma}

\begin{proof*}
	Let $t$ be produced by $(q,S)$ on input $s$ where $s\in T_\Sigma$.
	Clearly, it is sufficient to show that $\p{S}^A (t) \Rightarrow t $ due to Lemma~\ref{Property of A}.
	We prove our claim by structural induction.
	Let $s=a(s_1,\dots,s_k)$ where $a\in \Sigma_k$, $k\geq0$, and $s_1,\dots,s_k \in T_\Sigma$.
	As $t$ is produced by $(q,S)$ on input $s$, a right-hand side
	$\xi$ for $(q,S)$ and $a$ exists such that
	\begin{equation}\label{hatT_1 eq}
		t\in \xi [(q', S')(x_i) \leftarrow \p{(q',S')}_i^{\hat{T}_1} (s_i)
		\mid (q',S')\in \hat{Q}_1, i\in [k]].
	\end{equation}
	This means that
	\begin{equation*}
	t= \xi [u \leftarrow t/u \mid u\in V(\xi), \xi[u] \text{ is of the form } (q',S')(x_i)].
	\end{equation*}
	The following statements hold:
	\begin{enumerate}
		\item 	By definition of $\hat{T}_1$, it follows that $\p{S}^A(\xi) \Rightarrow \xi'$ where
		\[ \xi' = \xi [u\leftarrow   S'(u) \mid u\in V(\xi), \xi[u] \text{ is of the form } (q',S')(x_i) ] .\]
		\item  	Consider the node $u$. Let $u$ be labeled by $(q',S')(x_i)$ in $\xi$. 
		This means that $u$ is labeled by  $S'(u)$ in $\xi'$. 
		By  Equation~\ref{hatT_1 eq},
		$t/u$ can be produced by $(q',S')$ on input $s_i$.
		
		By induction hypothesis, 
		$\p{S'}^A (t') \Rightarrow t' $ for all trees $t'$ producible by $(q',S')$.
		Thus, $\p{S'}^A (t/u) \Rightarrow t/u$.
		By definition $\p{S'}^A (t/u) \Rightarrow t/u$ implies 
		\[
		\p{S'}^A_u (t/u) \Rightarrow t/u
		\]
		because $t/u$ is ground.
	\end{enumerate}
	Statements (1) and (2) yield that $\p{S}^A (t) \Rightarrow t $.
\end{proof*}

\noindent
We now show that the converse holds as well.

\begin{lemma}\label{T1' property 2}
	Let $s\in T_{\Sigma}$ and $t$ be producible by the state $q_1$ of $T_1$ on input $s$.
	Let $S \subseteq Q_2$ such that
	$t\in \bigcap_{q\in S} \text{dom}(q)$. Then $t$ be producible by the state $(q_1,S)$ of $\hat{T}_1$ on input $s$.
\end{lemma}

\begin{proof*}
	We prove our claim by structural induction.
	Let $s=a(s_1,\dots,s_k)$, $a\in \Sigma_k$, $k\geq 0$, and $s_1,\dots,s_k\in T_\Sigma$.
	As $t$ be producible by the state $q_1$ of $T_1$ on input $s$,
	$\xi \in \text{rhs}_{T_1} (q_1,a)$ exists such that
	\begin{equation*}
	t\in \xi [q'(x_{i}) \leftarrow \p{q'}^{T_1}_i (s_i) \mid q'\in Q_1, i\in [k]].
	\end{equation*}
	In essence, this means that
	\begin{equation*}
	t= \xi [u \leftarrow t/u \mid u\in V(\xi), \xi[u] \text{ is of the form } q'(x_i)].
	\end{equation*}
	Hence, it follows that if a node $u$ is labeled by $q'(x_i)$ in $\xi$ then
	\begin{equation}\tag{*}
		\p{q'}^{T_1}_i (s_i) \Rightarrow t/u.
	\end{equation}
	By our premise 	$t\in \bigcap_{q\in S} \text{dom}(q)$.
	Due to Lemma~\ref{Property of A}, it follows that $\p{S}^A (t)\Rightarrow t$.
	Therefore, for all leafs $u$ of $\xi$ 
	that are labeled by a symbol in of the form $q_1(x_i)$,
	a state $S_u$ of $A$ exists such that
	\[ \p{S}^A (\xi) \Rightarrow  \xi[u\leftarrow S_u (u) \mid u\in V(\xi), \xi[u]\in Q_1(X)] \]
	and 
	$\p{S_u}^A_u (t/u) \Rightarrow t/u$.
	By definition of $\hat{T}_1$, the former implies that  
	\[(q_1,S)(a(x_1,\dots,x_k))\rightarrow \xi[u\leftarrow (q',S_u) (x_i) \mid u\in V(\xi), \xi[u]=q'(x_i)]\ (\dagger) \] 
	is a rule of $\hat{T}_1$.
	The later implies $\p{S_u}^A (t/u) \Rightarrow t/u$ as $t_u$ is ground.
	Therefore, $t/u \in  \bigcap_{q\in S_u} \text{dom}(q)$ due to Lemma~\ref{Property of A}.
	
	Consider an arbitrary node $\breve{u}$ of $\xi$. Let $\breve{u}$ labeled by $\breve{q} (x_i)$ in $\xi$.
	Then,  $t/\breve{u} \in  \bigcap_{q\in S_{\breve{u}}} \text{dom}(q)$. 
	Furthermore, due to (*), it follows that 	$\p{\breve{q}}^{T_1}_i (s_i) \Rightarrow t/\breve{u}$.

	Then, the induction hypothesis yields that
	$\p{(\breve{q},S_{\breve{u}})}^{\hat{T}_1} (s_i) \rightarrow t/\breve{u}$.
	Note that $\p{(\breve{q},S_{\breve{u}})}^{\hat{T}_1} (s_i) \rightarrow t/\breve{u}$ implies $\p{(\breve{q},S_{\breve{u}})}^{\hat{T}_1}_i (s_i) \rightarrow t/\breve{u}$ because $s_i$ is ground.
	Along with $(\dagger)$, this yields our claim.
\end{proof*}

Lemmas~\ref{T1' property 1} and~\ref{T1' property 2} allow us to prove the following statement, which
implies Lemma~\ref{T1' property 4}.

\begin{lemma}\label{T1' property 3}
$\text{dom}(\hat{T}_1)=\text{dom}(T_1 \ci T_2)$ and 
for  $s\in T_\Sigma$,
$\hat{T}_1(s)=T_1(s) \cap \text{dom}(T_2)$.
\end{lemma}
\begin{proof*}
	First we show that $\text{dom}(\hat{T}_1)=\text{dom}(T_1 \ci T_2)$.
	Let $s\in \text{dom}(\hat{T}_1)$, i.e., a tree $t$ over $\Delta$ exists such that
	$\p{(q_1^0, \{q_2^0\})}^{\hat{T}_1} (s)\Rightarrow t$, where $(q_1^0, \{q_2^0\})$ is the initial
	state of $\hat{T}_1$.
	By construction of $\hat{T}_1$, it follows that $\p{q_1^0}^{{T}_1} (s)\Rightarrow t$ and by Lemma~\ref{T1' property 1}, $t\in\text{dom} (q_2^0)$. Hence $s\in \text{dom}(T_1 \ci T_2)$.
	For the converse, let $s\in \text{dom}(T_1 \ci T_2)$. Then, a tree $t$ over $\Delta$ exists such that 
	$\p{q_1^0}^{{T}_1} (s)\Rightarrow t$, where ${q_1^0}$ is the initial state of $T_1$,
	and $t\in\text{dom} (q_2)$. Hence, due to Lemma~\ref{T1' property 2} it follows that
	$\p{(q_1^0, \{q_2^0\})}^{\hat{T}_1} (s)\Rightarrow t$
	and thus, $s\in \text{dom}(\hat{T}_1)$.
	
	Now we show that  $\hat{T}_1(s)=T_1(s) \cap \text{dom}(T_2)$ .
	Let $\p{( q_1^0, \{q_2^0\})}^{\hat{T}_1} (s) \Rightarrow t$.
	By construction of $\hat{T}_1$, $\p{q_1^0}^{{T}_1} (s) \Rightarrow t$ holds.
	By Lemma~\ref{T1' property 1}, $t\in\text{dom} (q_2^0)$.
	Therefore, our claim follows.
	Conversely, let $t\in T_1(s) \cap \text{dom}(T_2)$. Then, clearly $t\in\text{dom}(q_2^0)$
	and $\p{q_1^0}^{{T}_1} (s) \Rightarrow t$.
	By Lemma~\ref{T1' property 2}, $\p{( q_1^0, \{q_2^0\})}^{\hat{T}_1} (s) \Rightarrow t$ which yields our claim.
\end{proof*}

%% file: appendix-iv.tex
\section{Correctness of the LA-Transducer $M$}
In this section, we present the formal proof of correctness for the la-transducer $M$, i.e.,
we show that $M$ is functional if and only if $T_1\ci T_2$ is functional.
Recall that due to Lemma~\ref{T1' property 4}, it is sufficient to consider
$\hat{T}_1\ci T_2$.

In the following, denote by $L$ the set of states of the la-automaton of $M$.
W.l.o.g. we assume that for all states  $l$ in $L$, $\text{dom}(l)\neq \emptyset$.
In the remainder of this section, our proofs employ partial trees in  $T_\Sigma [L]$.
Consider such a tree $s$. 
Recall that in a translation of $M$ input trees are first preprocessed by a 
relabeling induced by the la-automaton of $M$.
We demand that in a translation of $M$
the tree $s$ is relabeled as follows: If the $i$-th child of the node $v$ of $s$ is labeled by $l\in L$ then we require that $v$ be relabeled by a symbol of the form $\angl{a,l_1,\dots l_{i-1}, l, l_{i+1},\dots. l_k}$.

For instance, consider the la-automaton $B=(\{p,p'\},\Sigma,\Sigma,R,\{p\})$
where $\Sigma=\{f^2,a^0,b^0\}$ and $R$ contains the rules
\begin{center}
	$\begin{array}{lclclcl}
	p(f(x_1,x_2)) & \rightarrow & f(p(x_1), p(x_2)) & \quad & p(f(x_1,x_2)) & \rightarrow & f(p(x_1), p'(x_2))\\
	p'(f(x_1,x_2)) & \rightarrow & f(p'(x_1), p'(x_2)) & \quad & p'(f(x_1,x_2)) & \rightarrow & f(p'(x_1), p(x_2))\\
	p(a) & \rightarrow & a & \quad & p'(b) & \rightarrow & b.
	\end{array}$
\end{center}
Informally, the state $p$ checks whether or not the leftmost leaf of its input tree is $a$.
The state $p'$ does the same for $b$.
Consider the tree $s=f(a,f(p,b))$.
For $s$ the tree $\langle f,p,p\rangle(a ,\langle f,p,p'\rangle(p,b))$ is a valid relabeling.
The tree $\langle f,p,p'\rangle(a ,\langle f,p',p'\rangle(p,b))$ on the other hand is not.

\subsection{If $M$ is functional then $\hat{T}_1 \ci T_2$ is functional}
In this section we formally prove the only-if statement of  Corollary~\ref{main}, i.e., we show that
if $M$ is functional then $\hat{T}_1 \ci_2$ is functional. 
More precisely we formally show that $\mathcal{R} (\hat{T}_1) \circ \mathcal{R} (T_2) \subseteq \mathcal{R} (M)$. Obviously this implies our result.

In the following we formally prove Lemma~\ref{stronger subset claim}.
More precisely we prove the following lemma which is a more detailed version of Lemma~\ref{stronger subset claim}.

\begin{lemma}\label{lemma 1}
	Let $(q_1,S)$ be a state of $\hat{T}_1$ and $q_2$ be a state of $T_2$. 
	Let $s\in T_\Sigma$. Consider the state $(q_1,S,q_2)$ of $M$.
	If 
	\[
	\p{(q_1,S)}^{\hat{T}_1}(s) \Rightarrow t \quad\text{and}\quad\p{q_2}^{T_2} (t) \Rightarrow r
	\]
	then $\p{(q_1,S,q_2)}^M (s)\Rightarrow r$.
\end{lemma}

\begin{proof*}
	We prove our claim by induction on the structure of $s$.
	Let $s=a(s_1,\dots,s_k)$ where $a\in \Sigma_k$, $k\geq 0$, and for $i\in [k]$, $s_1,\dots,s_k\in T_\Sigma$.
	First, we prove the following claim.
	\begin{claim*}\label{claim} 	If 
		$\p{(q_1,S)}^{\hat{T}_1}(s) \Rightarrow t$ and $\p{q_2}^{T_2} (t) \Rightarrow r$ and $\p{(q_1,S,q_2)}^M (s) \Rightarrow r$, then
		trees $\xi$ and $\psi$  exist such that  
		\[
		\xi \in \text{rhs}_{\hat{T}_1}((q_1,S),a)\quad\text{and}\quad\p{q_2}^{T_2} (\xi) \Rightarrow \psi.\] 
		Furthermore, $\xi$ and $\psi$ have the following properties.
		\begin{enumerate}
			\item[(1)] Let $u$ be a node of $\xi$.
			If a node of $\psi$ is labeled by
			$q_2'(u)$  then a state $(q_1',S')$ of $\hat{T}_1$ exists such that
			$u$ is labeled by $(q_1',S')(x_\iota)$ in $\xi$, where $\iota\in [k]$, and $q_2'\in S'$.
			\item[(2)] It holds that 
			\[t\in \xi [q(x_i) \leftarrow \p{q}^{T_1'} (s_i) \mid q\in Q_1', i\in[k]]\]
			and
			\[r\in \psi [q(u) \leftarrow \p{q}^{T_2} (t/u)\mid q\in Q_2, u\in V(\xi) 
			].\]
		\end{enumerate}
	\end{claim*}	
	\begin{claimproof}
		By definition, $\p{(q_1,S)}^{\hat{T}_1}(s) \Rightarrow t$ implies $\p{q_1}^{{T}_1}(s) \Rightarrow t$ and thus
		\begin{equation}\tag{1}\label{equation}
		t\in \xi'[q(x_i) \leftarrow \p{q}^{T_1} (s_i) \mid q\in Q_1, i\in [k]]
		\end{equation}
		for some $\xi'\in \text{rhs} (q_1,a)$.
		Furthermore, by Lemma~\ref{T1' property 1}, $\p{(q_1,S)}^{\hat{T}_1}(s) \Rightarrow t$ implies that
		$t\in \bigcap_{q'\in S}\text{dom} (q')$.
		In the following, let $S=\{q_2^1,\dots,q_2^n\}$. 
		As $t\in \bigcap_{q'\in S}\text{dom} (q')$ and due to Equation~\ref{equation}, for all $j\in [n]$,  trees $\psi_j$ and $r_j$ exists such that 
		\begin{enumerate}
			\item[(a)] $\p{q_2^j}^{T_2} (\xi') \Rightarrow \psi_j$ and
			\item[(b)] $\p{q_2^j}^{T_2} (t) \Rightarrow r_j$ such that
			\[r_j\in \psi_j  [q_2'(u) \leftarrow \p{q_2'}^{T_2} (t/u)\mid q_2'\in Q_2, u\in V(\xi')].\]
		\end{enumerate}
		By our premise, the state $(q_1,S,q_2)$ of $M$ is defined which implies that $q_2\in S$. 
		W.l.o.g. let $q_2^1=q_2$. Furthermore, as $\p{q_2}^{T_2} (t) \Rightarrow r$, we can also assume
		that $r_1=r$.
		 
		Recall that by definition, if a node of $\psi_j$ is labeled by $q_2' (u)$, where $q_2'\in Q_2$ and
		$u$ is a node, then the node $u$ is labeled by some symbol in $Q_1(X)$ in $\xi'$.
	    Let $u_1,\dots, u_m$ be the nodes of $\xi'$ that are labeled by a symbol in $Q_1(X)$.
	    
	    In the following we first prove Statement~(1).
		Due to Lemma~\ref{A property 3}, (a) implies 
		\[\p{S}^A(\xi') \Rightarrow \xi' [u_i\leftarrow S_i (u_i)\mid i\in [m]],\]
		where $A$ is the domain automaton of $T_2$ and $S_i=\bigcup_{j\in [n]} \psi_j \s{u_i}$,
		which in turn implies that $\hat{T}_1$ contains the rule
		 $(q_1,S) (a(x_1,\dots,x_k)) \rightarrow \xi$ where
 		 \[\xi= \xi' [u_i\leftarrow (q',S_i) (x_\iota) \mid 
		 \xi[u_i]=q'(x_\iota), \iota\in [k]] \]
		as $\hat{T}_1$ is obtained from the p-construction of $T_1$ and $A$. 
		In the following, consider the node $u_i$. Assume that in  $\psi_j$ a node labeled by
		$q_2'(u_i)$ occurs. Recall that this means that the node $u_i$ is labeled by a symbol of the form
		$q'(x_\iota)$ in $\xi'$.
		By construction,  $u_i$ is labeled by $q'(x_\iota)$ in $\xi'$ if and only if
		$u_i$ is labeled by  $q'(x_\iota)$ in $\xi'$.  As $S_i=\bigcup_{j\in [n]} \psi_j \s{u_i}$, obviously $q_2'\in S_i$.

		As $q_2^1=q_2$, Statement~(1) follows with $\psi=\psi_1$.
		Note that clearly, for all $j\in [n]$, it holds that
		\begin{equation}\tag{2}\label{equation 2}
			\p{q_2^j}^{T_2} (\xi') \Rightarrow \psi_j\quad\text{if and only if}\quad \p{q_2^j}^{T_2} (\xi) \Rightarrow \psi_j.
		\end{equation}
		We now prove Statement~(2). 
		In particular, we prove the first part of Statement~(2), i.e., that
			\[t\in \xi [q(x_i) \leftarrow \p{q}^{T_1'} (s_i) \mid q\in Q_1', i\in[k]].\]
		Let the node $u_i$ be labeled by $(q'_1, S_i) (x_\iota)$ in $\xi$.
		Consider an arbitrary state $q_2'\in S_i$ where $i\in [m]$.
		In particular, this means that $q_2' \in \psi_j \s{u_i}$ for some $j\in [n]$.
		In other words, a node $g$ exists such that  $g$ is labeled by $q_2'(u_i)$ in $\psi_j$.
		Clearly, Statement (b) implies that
		$q_2'$ can produce  the tree $r_j/g$ on input $t/u_i$. This statement can be generalized.
		More precisely, it holds that any state $q_2'\in S_i$ can produce some output tree on input
		$t/u_i$.
		Therefore, $t/u_i \in  \bigcap_{q_2'\in S_i}\text{dom} (q_2')$. 
		Equation~\ref{equation} implies that $\p{q'_1}^{T_1}(s_\iota) \Rightarrow t/u_i$ if $\xi'[u_i]=q'_1(x_\iota)$.
		Together with Lemma~\ref{T1' property 2} and as $t/u_i \in  \bigcap_{q_2'\in S_i}\text{dom} (q_2')$, it follows that
		\[
		\p{(q'_1, S_i)}^{\hat{T}_1} (s_\iota) \Rightarrow t/u_i.
		\]
		By construction the node $u_i$ is labeled by $(q'_1, S_i) (x_\iota)$ in $\xi$ if and only if
		$u_i$ is labeled by $q' (x_\iota)$ in $\xi'$.
		With the rule $(q_1,S) (a(x_1,\dots,x_k)) \rightarrow \xi$ and Equation~\ref{equation}
		it follows that the $(q_1,S)$
		can generate the tree $t$ on input $s$. More precisely, it follows that
		\[t\in \xi [q(x_i) \leftarrow \p{q}^{\hat{T}_1} (s_i) \mid q\in Q_1', i\in[k]].\]
		The second part of Statement~(2) follows due to Statement~(b) and Equation~\ref{equation 2}.
	\end{claimproof}
	Let $\xi$ and $\psi$ be as in Claim~\ref{claim}. Due to Statement~(1) of Claim~\ref{claim},
	it follows that $M$ contains the rule 
	\[(q_1, S, q_2) (a(x_1  \!:\! l_1,\dots,x_k  \!:\! l_k)) \rightarrow \gamma\] where 
	$\gamma$ is obtained from $\psi$ by substituting occurrences of $q_2'(u)$ in
	$\psi$, where $q_2' \in Q_2$ and $u$ is a leaf of $\xi$ labeled by a symbol of the form
	$(q'_1,S') (x_i)$, by $(q_1',S',q_2') (x_i)$.
	Furthermore,  for $i\in [k]$,
	$\xi \s{x_i}=l_i$.
	
	We now show that  $\p{(q_1,S,q_2)}^M (s)\Rightarrow r$. Recall that $s=a(s_1,\dots,s_k)$ where $a\in \Sigma_k$, $k\geq 0$, and for $i\in [k]$, $s_1,\dots,s_k\in T_\Sigma$.
	Note that as $\xi \s{x_i}=l_i$ and due to
	Statement~(2) of Claim~\ref{claim} and Lemma~\ref{Property of A}, $s_i\in \text{dom} (l_i)$ for $i\in [k]$.
	
	Consider a node $g$. By definition of $\gamma$, $g$ is labeled by  $(q_1',S',q_2') (x_i)$ in $\gamma$ 
	if and only if
	$g$ is labeled by $q'_2 (u)$ in $\psi$ and the node $u$ is labeled by
	$(q_1',S') (x_i)$ in $\xi$.
	Statement~(2) of Claim~\ref{claim} implies that
	$\p{(q_1',S')}^{\hat{T}_1} (s_i)\Rightarrow t/u$ and $\p{q'_2}^{T_2} (t/u)\Rightarrow r/g$.
	Therefore, by induction hypothesis,  $\p{(q_1',S',q_2')}^{M} (s_i)\Rightarrow r/g$.
    Clearly, our claim follows.
\end{proof*}

\noindent
Clearly, Lemma~\ref{lemma 1} implies Lemma~\ref{subset}. Lemma~\ref{lemma 1} also yields the following two
auxiliary results.

\begin{lemma}\label{M dom lemma}
	Let $(q_1,S)$ be a state of $\hat{T}_1$ and  $q_2$ be a state of $T_2$
	such that $q_2\in S$. Then, for the state $(q_1,S,q_2)$ of $M$, 
	$\text{dom} ((q_1,S,q_2)) = \text{dom} ((q_1,S))$ holds.
\end{lemma}
\begin{proof*}
	Let $s=a(s_1,\dots,s_k)$, $a\in \Sigma_k$, $k\geq 0$ and $s_1,\dots,s_k\in T_\Sigma$.
	Let $s\in \text{dom} ((q_1,S,q_2))$, i.e., $\p{(q_1, S, q_2)} (s) \Rightarrow r$ for some tree $r$.
	Consider the first rule of $M$ applied in this translation. Let
	\[  \eta=(q_1,S,q_2) (a (x_1\!:l_1,\dots, x_k\!:l_k ) ) \rightarrow \gamma \]
	be this rule. By construction $\eta$ is obtained from a rule
	$(q_1,S) (a (x_1,\dots, x_k) ) \rightarrow \xi$
	of $\hat{T}_1$ such that for $i\in [k]$, $\xi \s{x_i} \subseteq l_i$.
	The application of $\eta$ implies $s_i\in l_i$ for $i\in [k]$.
	This implies  $s\in \text{dom} ((q_1, S))$.
	
	Conversely, let $s\in \text{dom} ((q_1, S))$, i.e., $\p{(q_1,S)} (s) \rightarrow t$
	for some tree $t$. 
	Note that the state $(q_1,S,q_2)$ of $M$ implies $q_2\in S$.
	Due to Lemma~\ref{T1' property 1}, it follows that $ \p{q_2}^{T_2} (t) \neq \emptyset$. Therefore, we deduce 
	that  due to  Lemma~\ref{lemma 1}, $s\in \text{dom} ((q_1,S,q_2))$.
\end{proof*}

\begin{lemma}\label{lemma aux}
	Let $s\in T_\Sigma [L]$. Let
	$M(s) \Rightarrow r_M$ and let $(q_1,S,q_2) (v)$ occurs in $r_M$, where
	$(q_1,S,q_2)$ is a state of $M$ and $v$ is a node of $s$ labeled by a symbol
	$l\in L$.
	Then $\text{dom} (l) \subseteq \text{dom} ((q_1,S,q_2))$. 
\end{lemma}
\begin{proof*}
Let the parent node of $v$ be labeled by $a\in \Sigma_k$ where $k>0$.
W.l.o.g. let $v$ be the first child of its parent node. 
Then, clearly the occurrence of $(q_1,S,q_2) (v)$ in $r_M$ originates from the application of a rule
$(q_1',S',q_2') (a(x_1\!:l_1,\dots, x_k\!:l_k))\rightarrow \gamma$ of $M$ such that
$(q_1,S,q_2) (x_1)$ occurs in $\gamma$. Recall that by definition,
$l_1,\dots l_k$ are sets of states of $\hat{T}_1$.
By the  definition of relabelings of trees in $T_\Sigma [L]$, the parent node
 of $v$ is relabeled by a symbol of the form
$\angl{a,l, l'_2,\dots l_k'}$ 
which implies  $l=l_1$.

Consider the rule $(q_1',S',q_2') (a(x_1\!:l_1,\dots, x_k\!:l_k))\rightarrow \gamma$.
Recall that by construction of $M$, this rule is obtained from a rule
$(q_1',S') (a(x_1,\dots, x_k))\rightarrow \xi$ of $\hat{T}_1$ such that for $i\in [k]$,
$\xi \s{x_i} \subseteq l_i$.  Note that the occurrence of $(q_1,S,q_2) (x_1)$ in $\gamma$
implies that $(q_1,S) (x_1)$ occurs in $\xi$.
Therefore, the state $(q_1,S)$ of $\hat{T}_1$ is included in $l$.
As $s\in \text{dom} (l)$ if and only if $s\in\bigcap_{\hat{q}\in l} \text{dom} (\hat{q})$, our claim follows due to Lemma~\ref{M dom lemma}.
\end{proof*}

\subsection{If $\hat{T}_1 \ci T_2$ is functional then $M$ is functional.}
In this section we formally prove the only-if statement of  Corollary~\ref{main}, i.e., we show that
if $\hat{T}_1 \ci T_2$ is functional then $M$ is functional. 

First we introduce the following definition.
Recall that we have introduced \emph{synchronized} translations of $M$ in
Section~\ref{construction}.
In the following, we extend this definition.
Let $s\in T_\Sigma [L]$.
We call the trees $s$, $t$, $r$ and $r_M$  \emph{synchronized} if
\begin{enumerate}
\item $\hat{T}_1 (s) \Rightarrow t$ and $T_2(t) \Rightarrow r$ and $M(s) \Rightarrow r_M$ and
\item the tree $r_M$ is obtained from $r$ by substituting all occurrences of $q_2'(u)$ in
$r$ by $(q_1',S',q_2') (v)$,
where $(q'_1,S')$ and $q_2'$ are states of $\hat{T}_1$ and $T_2$, respectively,
 and $u$ is a leaf of $t$ labeled by $(q'_1,S') (v)$.
\end{enumerate}

Informally, $s$, $t$, $r$ and $r_M$ are \emph{synchronized} if
on input $s$, $M$ produces the tree $r_M$ by accurately
simulating $\hat{T}_1 \ci T_2$.
More precisely:
Recall that when a state $(q_1,S,q_2)$ of $M$ processes a subtree $s'$ of $s$
then $(q_1,S,q_2)$ guesses what the state $(q_1,S)$ of $\hat{T}_1$ might have produced
before producing output according to this guess. 
Informally, if all such guesses of $M$ are correct, i.e., the states of $\hat{T}_1$
have indeed produced the trees $M$ has guessed, then 
$s$, $t$, $r$ and $r_M$ are \emph{synchronized}.

Before we prove a more detailed version of Lemma~\ref{aux},
recall that by definition, a state $l$ in $L$ is a set of states of $\hat{T}_1$.
Consider a tree $s\in T_\Sigma [L]$. Informally, if a symbol $l\in L$ occurs at
some leaf of $s$ then $l$ can be considered a placeholder for some tree $s'$ such that
$s'\in \bigcap_{(q_1,S) \in l} \text{dom} (q_1,S)$.
We now show that the following holds.

\begin{lemma}\label{lemma 2}
	Let $s\in T_\Sigma [L]$. Let
	$M(s)\Rightarrow r_M$ and let $(q_1,S,q_2)(v)$ occur in $r_M$. Then trees
	$t$, $r$ and $r_M'$ exist such that $s$,  $t$, $r$ and $r_m'$ are synchronized and 
	$(q_1,S,q_2)(v)$ occurs in $r'_M$.
	Furthermore, let $v'$ be a leaf of $s$ that is labeled by $l\in L$.
	Then,
	it holds that
	if  $(q'_1,S')(v')$ occurs in $t$ then $(q'_1,S')\in l$.
\end{lemma}

\begin{proof*}
	We prove our claim by structural induction. 
	First, let $\bar{v}$ be a node of $s$ 
	such that the subtree of $s$ rooted at $\bar{v}$ is of the form
	 $a(l_1,\dots, l_k)$ where $a\in \Sigma_k$, $k\geq 0$, and $l_1,\dots,l_k \in L$.
	 Note that by definition $\bar{v}$ can be a leaf.
	Then, a state $l\in L$ exists such that
	\[
	l(a(x_1,\dots,x_k))\rightarrow  a(l_1 (x_1),\dots, l_k (x_k))
	\]
	is a rule of the la-automaton of $M$.
	Furthermore, as $M(s)\Rightarrow r_M$, on input $\bar{s}=s[\bar{v}\leftarrow l]$, the 
	$M$ produces 
	the tree $\bar{r}_M$ such that
	\begin{equation*}
		r_M \in  \bar{r}_M [(q)(\bar{v}) \leftarrow \p{q}^M_{\bar{v}} (a(l_1,\dots, l_k)) \mid q\text{ is a state of } M]. 
	\end{equation*}
	We remark that all trees in $\p{q}^M_{\bar{v}} (a(l_1,\dots, l_k))$ are of the form
	$\gamma\langle x_i\leftarrow \bar{v}.i\mid i\in [k]\rangle$ where
	$\gamma \in \text{rhs} (q,a,l_1,\dots,l_k)$. 
	Recall that by our premise,  $(q_1,S,q_2)(v)$ occurs in $r_M$.
	Then one of the following cases arises:
	\begin{enumerate}
		\item[(a)] $(q_1,S,q_2)(v)$ does not already occur in  $\bar{r}_M$.
		\item[(b)] $(q_1,S,q_2)(v)$ already occurs in  $\bar{r}_M$.
	\end{enumerate}
	First, we consider case (b). 
	By induction hypothesis, as $M (\bar{s}) \Rightarrow \bar{r}_M$ and a node labeled by $(q_1,S,q_2) (v)$ occurs in $\bar{r}_M$,
	trees $\bar{t}$, $\bar{r}$ and $\bar{r}_M'$ exist such that $\bar{s}$,  
	$\bar{t}$, $\bar{r}$ and $\bar{r}_M'$ are synchronized and 
	$(q_1,S,q_2) (v)$ occurs in $\bar{r}'_M$.
	Furthermore, by induction hypothesis, it holds that if  $(q'_1,S')(\bar{v})$ occurs in $\bar{t}$
	then $(q'_1,S') \in l$. 
	
	First, we construct the tree $t$.
	Recall that the la-automaton of $M$ is the domain automaton of $\hat{T}_1$.
    Therefore the existence of rule
    \[
    l(a(x_1,\dots,x_k))\rightarrow  a(l_1 (x_1),\dots, l_k (x_k))
    \]
    of the la-automaton implies
    that for all states $(q'_1,S') \in l$, a right-hand side
    $\xi' \in \text{rhs}_{\hat{T}_1} ((q_1',S'),a)$ exists such that   $\xi'\s{x_i}\subseteq l_i$ for $i\in [k]$ $(*)$.

	In the following, we define $t_{(q_1',S')}= \xi'\langle x_i\leftarrow  \bar{v}.i\mid i\in [k]\rangle$ if $(q'_1,S') \in l$. Then
	clearly $\hat{T}_1 (s)\Rightarrow t$ where
	\begin{equation*}
		t=\bar{t} [(q'_1,S')(\bar{v}) \leftarrow t_{(q_1',S')} \mid (q'_1,S') \in Q_1']. 
	\end{equation*}
	We now show that for arbitrary nodes $v'$ of $s$ it holds that
	if $v'$ is labeled by $l$ in $s$ and
	$(q'_1,S')(v')$ occurs in $t$ then $(q'_1,S')\in l$.
	Due to (*), this holds for all nodes $v'$ that are descendants of  $\bar{v}$.
	Now assume that $v'$ is not be a descendant of $\bar{v}$. 
	Let $v'$ be labeled by the symbol $\bar{l} \in L$ in $s$.
	Then obviously, the node $v'$ is also labeled by $\bar{l}$ in $\bar{s}$.
	Thus, by definition of $\bar{t}$, if $(q'_1,S')(v')$ occurs in $\bar{t}$ then $(q'_1,S')\in \bar{l}$.
	By construction of $t$, $(q'_1,S')(v')$ occurs in $t$ if and only if $(q'_1,S')(v')$ occurs in $\bar{t}$.
	This yields our claim.

	We now construct $r$ and $r_M'$. First recall that, by induction hypothesis, the
	trees $\bar{s}$,  
	$\bar{t}$, $\bar{r}$ and $\bar{r}_M'$ are synchronized.
	Therefore, for an arbitrary node $g$  the following holds:
	 $g$ is labeled by $(q'_1,S',q_2') (v)$
	in $\bar{r}_M$
	if and only if $g$ is labeled by $q_2' (u)$ in $\bar{r}$ and $u$ is a node of 
	$\bar{t}$ labeled by $(q'_1,S') (v)$ $(\dagger)$.

	Now let the node $g$  be labeled by $q_2' (u)$ in $\bar{r}$ and  let the node $u$  be
	labeled by   $(q'_1,S') (\bar{v})$ in $\bar{t}$. 
	Consider the right-hand side $\xi'$  
	assigned to the state $(q'_1,S')$ in $(*)$.
	Due to how rules of $\hat{T}_1$ are defined, it holds that
	\[
	\p{S'}^A(\xi')\Rightarrow \xi'[u\leftarrow \bar{S}(u) \mid u\in V(\xi'), \xi'[u]=(\bar{q},\bar{S})(x_i) ]. 
	\]
	Note that $(\dagger)$ implies $q_2'\in S$. This follows as the state $(q'_1,S',q_2')$
	is defined.
	Therefore by Lemma~\ref{A property 4}, a tree $\psi'$ exists such that 
	\begin{enumerate}
		\item  $\p{q_2'}^{T_2} (\xi') \Rightarrow \psi'$ and
		\item  if the node $u'$ is labeled by $(\tilde{q},\tilde{S})(x_i)$ in $\xi'$ then ${\psi'}\s{u'} \subseteq \tilde{S}$. 
	\end{enumerate}
	The later implies that if $\tilde{q}_2(u')$ occurs in $\psi'$ then $\tilde{q}_2\in \tilde{S}$.
	Due to $(*)$, for $i\in [k]$, it holds that  $\xi'\s{x_i}\subseteq l_i$.
	Therefore, by construction of $M$ the rule
	 \[
	 (q'_1,S',q_2')  (a (x_1\!:l_1,\dots, x_k\!: l_k)) \rightarrow \gamma'
	 \] 
	is defined where
	$\gamma'$ is obtained from $\psi'$ by substituting occurrences of $\tilde{q}_2(u')$ in
	$\psi$  by $(\tilde{q}_1,\tilde{S},\tilde{q}_2) (x_i)$,
	where $(\tilde{q}_1,\tilde{S})$ and $\tilde{q}_2$ are states of $\hat{T}_1$ and $T_2$, respectively,
	and $u'$ is a leaf of $\xi'$ labeled by a symbol of the form
	$(\tilde{q}_1,\tilde{S}) (x_i)$.
	
	For the node $g$ we define
	$r_{T_2,g} = \psi'\langle u'\leftarrow u.u' \mid u'\in V \rangle$.
	Additionally, we define
	$r_{M,g} =\gamma'\langle x_i\leftarrow \bar{v}.i\mid i\in [k] \rangle$.
	
	Recall that $(\dagger)$ holds.
	Then, $T_2(t)\Rightarrow r$ where
	\[
	r=\bar{r}  [g\leftarrow r_{T_2, g} \mid 
	\bar{r}[g] =q_2' (u) \text{ and }
	\bar{t}[u]=(q_1',S') (\bar{v})] 
	\]
	and $M(s)\Rightarrow r_M'$ where
	\[
	r_M'=\bar{r}_M'[  g \leftarrow r_{M,g}  \mid \bar{r}_M [g] = (q_1', S', q_2')(\bar{v})].
	\]
	Note that the node $\bar{v}$ of $s$ is relabeled by $\angl{a,l_1,\dots,l_k}$ via the relabeling induced by the rule $l(a(x_1,\dots,x_k))\rightarrow  a(l_1 (x_1),\dots, l_k (x_k))$ of the la-automaton of $M$.
	Thus, $r_M'$ is well defined.
	Clearly, $\hat{T}_1 (s) \Rightarrow t$ and $T_2(t) \Rightarrow r$ and $M(s) \Rightarrow r_M'$.
	Due to $(\dagger)$ and the construction of $r$ and $r_M'$, it follows that the second part of the synchronized-property holds as well.\\
	
	We now consider case (a).
	As  $(q_1,S,q_2)(v)$ occurs in $r_M$ but not in $\bar{r}_M$, it follows that
	$v=\bar{v}.i$ for some $i\in [k]$. 
	W.l.o.g. let  $v=\bar{v}.1$, i.e., $v$ is the first child of the node $\bar{v}$.
	Furthermore, it follows that a rule
	\[
	(\tilde{q}_1,\tilde{S},\tilde{q}_2) (a(x_1\!:\!l_1,\dots,x_k\!:\!l_k ))\rightarrow \tilde{\gamma} \]
	exists
	such that 	$(\tilde{q}_1,\tilde{S},\tilde{q}_2) (\bar{v})$
	occurs in $\bar{r}_M$  and $(q_1,S,q_2)(x_1)$ occurs in $\tilde{\gamma}$.
	Let the rule of $M$ above be obtained from the rule
	$(\tilde{q}_1,\tilde{S})(a(x_1,\dots,x_k))\rightarrow \tilde{\xi}$
	of $\hat{T}_1$ and subsequently 
	translating $\tilde{\xi}$ by the state $\tilde{q}_2$ of $T_2$.
	In particular, this means that a tree $\tilde{\psi}$ exists such that 
	\begin{enumerate}
		\item[(a)] $\p{\tilde{q}_2}^{T_2}  (\tilde{\xi})\Rightarrow \tilde{\psi}$ and
		\item[(b)]   $\tilde{\gamma}$ is obtained from $\tilde{\psi}$ by substituting occurrences of $q_2'(u)$ in
		$\tilde{\psi}$ by $(q_1',S',q_2') (x_i)$, where $(q_1',S')$ and
		$q_2'$ are states of $\hat{T}_1$ and $T_2$, respectively, 
		and $u$ is a leaf of $\xi'$ labeled by  $(q'_1,S') (x_i)$, 
	\end{enumerate}

	By induction hypothesis, 
	as $M (\bar{s}) \Rightarrow \bar{r}_M$ and a node labeled by $(\tilde{q}_1,\tilde{S},\tilde{q}_2) (\bar{v})$ occurs in $\bar{r}_M$,
	trees $\bar{t}$, $\bar{r}$ and $\bar{r}_M'$ exist such that $\bar{s}$,  
	$\bar{t}$, $\bar{r}$ and $\bar{r}_M'$ are synchronized and 
	$(\tilde{q}_1,\tilde{S},\tilde{q}_2) (\bar{v})$ occurs in $\bar{r}'_M$.

	Let the node $\bar{g}$ be labeled by $(\tilde{q}_1,\tilde{S},\tilde{q}_2) (\bar{v})$ 
	in $\bar{r}'_M$.
	Due to the synchronized property, the node $\bar{g}$ is labeled by
	$q_2 (\bar{u})$ in $\bar{r}$, where $\bar{u}$
	is a node  that is labeled by $(q_1,S)(\bar{v})$ in $\bar{t}$. 
	
	To construct the trees $t$, $r$ and $r_M'$,
	we then proceed as in case (b) but set
	\begin{itemize}
		\item $t_{(\tilde{q}_1,\tilde{S})}= \tilde{\xi}\langle x_i\leftarrow  \bar{v}.i\mid i\in [k]\rangle$,
		\item $r_{T_2,\hat{g}} = \tilde{\psi}\langle u'\leftarrow \bar{u}.u' \mid u'\in V \rangle$
		and 
		\item $r_{M,\hat{g}} = \tilde{\gamma}\langle x_i\leftarrow \bar{v}.i\mid i\in [k] \rangle$.
	\end{itemize}
	This  yields our claim.
\end{proof*}

Lemma~\ref{lemma 2} and  and Proposition~\ref{prop} allow us to formally prove the following
version of Lemma~\ref{main lemma}.

\begin{lemma}\label{lemma 5}
	Let $s\in T_\Sigma [L]$ such that
	only a single node $v$ of $s$ is labeled by a symbol in $L$.
	Let $v$ be labeled by $l \in L$.
	Let
	$M(s)\Rightarrow r_M$ such that $(q_1,S,q_2)(v)$ occurs in $r_M$.
	
	Consider the tree $\tilde{s}= s[v\leftarrow s']$ where $s'\in \text{dom} (l)$. 
	If $T_1\ci T_2(\tilde{s})$ is a singleton then 
	$\p{(q_1,S,q_2)} (s')$ is a singleton. 
\end{lemma}

\begin{proof*}
	Note that by Lemma~\ref{lemma aux},
	$s' \in \text{dom} ((q_1,S,q_2))$. Hence, 
	 $\p{(q_1,S,q_2)} (s') \neq \emptyset$.
	Assume that $\p{(q_1,S,q_2)} (s')$ is not a singleton, i.e., assume that
	distinct trees $r_1$, $r_2$ exist such that $r_1,r_2\in \p{(q_1,S,q_2)} (s')$. 

	We claim that for $r_1$, a tree $t_1$ exists such that
	\begin{enumerate}
	\item on input $s'$, the state $(q_1,S)$ of $\hat{T}_1$ produces $t_1$ and
	\item on input $t_1$, the state $q_2$ of $T_2$ produces $r_1$.
	\end{enumerate}
	We will later prove this claim in detail.
	It can be shown that a tree $t_2$ with the same properties exists for $r_2$.
   
	Using this claim and Proposition~\ref{prop}, we now prove that contrary to the assumption $r_1= r_2$.
	
	Due to Lemma~\ref{lemma 2}, as 	$M(s)\Rightarrow r_M$ and $(q_1,S,q_2)(v)$ occurs in $r_M$,
	it follows that trees  $t$, $r$ and $r_M'$ exist
	such that $s$,  $t$, $r$ and $r_M'$ are synchronized and 
	$(q_1,S,q_2)(v)$ occurs in $r_M'$.  Moreover, 
	if $(q',S')(v)$ occurs in $t$, where $(q',S')$ is some state of $\hat{T}_1$,
	then $(q',S')\in l$. Recall that by our premise, $v$ is labeled by $l$ in $s$.
	 Therefore, $\text{dom} (l) \subseteq \text{dom} ( (q',S') )$ due to Lemma~\ref{Property of A}.
	Consequently,  $s' \in \text{dom} ((q',S'))$.
	Therefore, for all states $(q',S')$ of $\hat{T}_1$ such that $(q',S') (v)$ occurs in $t$,
	a tree $t'$ exists such that $\p{(q',S')}^{\hat{T}_1}(s') \Rightarrow t'$.
	In the following, let $t\s{v}= \{ (q^1,S^1),\dots, (q^n,S^n) \}$ and
	for $j\in [n]$, let $\p{ (q^j,S^j) }^{\hat{T}_1} (s')\Rightarrow t_j'$.
	Then clearly on input $\tilde{s}$, the transducer
	$\hat{T}_1$  can produce the tree $\tilde{t}$ where 
		\[
		\tilde{t}= t[(q^j,S^j) (v)\leftarrow t_{j}' \mid j\in [n] ].
		\]
	Now consider the tree $r$. Let $q'_2$ be a state of $T_2$ and $u$ be a node.
	By definition of $t$ and $r$, if $q'_2(u)$ occurs in $r$, then the node $u$ is labeled by a symbol of the form $(q',S')(v)$ in $t$. Furthermore, the synchronized property implies that
	$q_2' \in S'$. This follows as a state $(q',S',q_2')$ of $M$ has the property that
	$q_2'\in S'$.
	The subtree of $\tilde{t}$ rooted at $u$
	is a tree $t'$ such that $\p{(q',S')}^{\hat{T}_1}(s') \Rightarrow t'$. By Lemma~\ref{T1' property 1},
	it follows that $t' \in \text{dom} (q_2)$.
	Therefore, it follows easily that
	$T_2(\tilde{t})\Rightarrow \tilde{r}$ where
	\[
	\tilde{r}=r[ q_2' (u) \leftarrow r_u \mid  q_2'\in Q_2, \tilde{t}[u]=t_j'\text{ and }  \p{q_2'}^{T_2}(t_j')\Rightarrow r_u  ].
	\]
	
	By our premise a node $g$ exists such that $g$ is labeled by $(q_1,S,q_2)(v)$ in 
	$r_M'$.  As the trees $s$, $t$, $r$ and $r_M'$ are synchronized,
	$g$ is labeled by $q_2(u)$  in $r$ where $u$ is a node of $t$ such that
	$t[u]=(q_1,S)(v)$.
	Due to our claim, a tree $t_1$ exists such that	$\p{(q_1,S)}^{\hat{T}_1} (s') \Rightarrow t_1$ and
	and $\p{q_2}^{T_2} (t_1) \Rightarrow r_1$.
	W.l.o.g. we  assume that $(q^1,S^1)=(q_1,S)$ and $t_1'=t_1$.
	Then, it follows easily that on input $\tilde{s}$, the composition $\hat{T}_1 \ci T_2$
	can produce a tree $\tilde{r}_1$ such that $\tilde{r}_1/g =r_1$.
	Analogously, it follows easily that on input $\tilde{s}$, the composition $\hat{T}_1 \ci T_2$
	can produce a tree $\tilde{r}_2$ such that $\tilde{r}_2/g =r_2$.
	Due to Proposition~\ref{prop},
	$\tilde{r}_1=\tilde{r}_2$ and therefore \[r_1=\tilde{r}_1/g=\tilde{r}_2/g=r_2.\]
	
	Now all that is left is to prove our previous claim that for $r_1$ and $r_2$, trees $t_1$
	and $t_2$ exist such that 
	\begin{itemize}
		\item   $\p{(q_1,S)}^{\hat{T}_1} (s') \Rightarrow t_1$ and
		and $\p{q_2}^{T_2} (t_1) \Rightarrow r_1$
		\item   $\p{(q_1,S)}^{\hat{T}_1} (s') \Rightarrow t_2$ and
		and $\p{q_2}^{T_2} (t_2) \Rightarrow r_2$.
	\end{itemize}
	We prove our claim for $r_1$. The proof for $r_2$ is analogous. 
	Let $s'= a(s_1,\dots,s_k)$ where $a\in \Sigma_k$, $k\geq 0$, and $s_1,\dots,s_k\in T_\Sigma$.
	As $r_1$ is producible by ${(q_1,S,q_2)}$ on input  $s'$, it follows that
	\begin{equation}\label{r1 equation}
	r_1 \in \gamma [ q_M(x_i) \leftarrow \p{q_M} (s_i) \mid i\in [k]\text{ and } q_M \text{ is a state of } M]
	\end{equation}
	where $(q_1,S,q_2) (a (x_1\!:l_1,\dots,x_k\!:l_k))\rightarrow \gamma$ is a rule of $M$, $l_1,\dots, l_k$ are
	states of the la-automaton of $M$ and for $i\in [k]$,
	 $s_i\in \text{dom} (l_i)$. 
	 
	 Before we prove our claim, we prove the following result  by induction on the statement of
	 Lemma~\ref{lemma 5}.
	
	\begin{claim*}\label{lemma 5 claim 1}
		Let $q_M$ be a state of $M$ and let $q_M(x_i)$ occur $\gamma$ where $i\in [k]$. Then the set $\p{q_M} (s_i)$ is a singleton.
	\end{claim*}
	\begin{claimproof}		
		Before, we prove our claim  consider the following.
		Let $q_M'$ be a state of $M$.
		Then, by Lemma~\ref{lemma aux},
		$s' \in \text{dom} (q_M')$, if $q_M' (v)$ occurs in $r_M$~$(\dagger)$.
		
		Now, we prove our claim.
		W.l.o.g., we consider the case $i=1$.
		Consider the tree $\bar{s}=a(l_1,s_2,\dots,s_k)$. 
		Due to $(\dagger)$,
		it follows that on input $\bar{s}$, any state $q_M'$ such that $q_M' (v)$ occurs in $r_M$
		 can produce some partial tree, i.e.,  a tree with leafs
		with label of the form $\breve{q}_M(v.1)$ where $\breve{q}_M$ is a state of $M$.
		In particular, by our premise, $(q_1,S,q_2)(v)$ occurs in $r_M$.
		As the tree  $r_1$ is producible by  $(q_1,S,q_2)$ on input  $s'$ by applying the rule
		 $(q_1,S,q_2) (a (x_1\!:l_1,\dots,x_k\!:l_k))\rightarrow \gamma$,
		it follows easily that on input $\bar{s}$, the state $(q_1,S,q_2)$ generates a tree $\hat{r}$ such
		that $q_M(v.1)$ occurs in $\hat{r}$ if $q_M(x_1)$ occurs $\gamma$.

		Thus, it follows that $M$ on input $s[v\leftarrow \bar{s}]$ produces a tree in which  $q_M(v.1)$ occurs. Clearly,
		the node $v.1$ is labeled by $l_1$ in $\bar{s}$.
		 Note that $s_1\in \text{dom} (l_1)$. By induction hypotheses, $\p{q_M} (s_i)$ is a singleton.
	\end{claimproof}

	\noindent
	We now prove our main claim.
	
	\begin{claim*}\label{lemma 5 claim 2}
		A tree $t_1$ exists such that  $\p{(q_1,S)}^{\hat{T}_1} (s') \Rightarrow t_1$ and
		and $\p{q_2}^{T_2} (t_1) \Rightarrow r_1$.
	\end{claim*}
	\begin{claimproof}
		Let $i\in[k]$ and $q_M$ be a state of $M$.
		Let $q_M (x_i)$ occur in $\gamma$.
		By Claim~\ref{lemma 5 claim 1}, the set $\p{q_M}(s_i)$
		is a singleton. Let $\p{q_M} (s_i)=\{r'\}$. Let $q_M=(q'_1,S',q'_2)$
		where $(q_1',S')$ and $q_2'$ are states of $\hat{T}_1$ and $T_2$, respectively.
		Due to Lemma~\ref{M dom lemma},
		$s_i\in \text{dom}(q_M)$ implies $s_i \in \text{dom} ((q'_1, S'))$, i.e.,
		$\p{(q_1',S')}^{\hat{T}_1} (s_i)$ is not empty.
		In the following, we show that 
		if $\p{q_M} (s_i)=\{r'\}$ then
		for all trees $t'$ contained in
		$\p{(q_1',S')}^{\hat{T}_1} (s_i)$,
		it holds that $\p{q_2'}^{T_2} (t')=\{r'\}$ $(*)$.
		
		In the following, consider such a tree $t'$.
		By Lemma~\ref{T1' property 1}, 
		\[
		t' \in \bigcap_{\bar{q}_2 \in S'}\text{dom} (\bar{q}_2)
		\] 
		and thus $t' \in \text{dom}(q'_2)$. Recall that by definition the state $q_M=(q'_1,S',q'_2)$ implies $q_2'\in S'$.
		Therefore, the set $\p{q_2'}^{T_2} (t')$ is not empty.
		As $\p{q_M} (s_i)=\{r'\}$, we deduce that
		$\p{q_2'}^{T_2} (t')=\{r'\}$ due to Lemma~\ref{lemma 1}.
		Therefore, $(*)$ follows. 
		
		By definition, the rule $(q_1,S,q_2) (a (x_1\!:l_1,\dots,x_k\!:l_k))\rightarrow \gamma$ of $M$ is defined only if a rule $(q_1,S)(a(x_1,\dots,x_k)) \rightarrow \xi$ of $\hat{T}_1$ and a tree $\psi$ exist such that
		\begin{enumerate}
			\item 	$\p{q_2}^{T_2} (\xi)\Rightarrow \psi$
			\item 	the tree $\gamma$ is obtained from $\psi$ by substituting all occurrences of $q_2'(u)$ in
			$\psi$ by $(q_1',S',q_2') (x_i)$,
			 where $(q'_1,S')$ and $q_2'$ are states of $\hat{T}_1$ and $T_2$, respectively, 
			 and  $u$ is a leaf of $\xi$ labeled by 
			$(q'_1,S') (x_i)$
			\item for $i\in [k]$, it holds that  $\xi \s{x_i}\subseteq l_i$.
		\end{enumerate}
		By definition of $r_1$ (see~Equation~\ref{r1 equation}),  for $i\in [k]$, it holds that 
		$s_i\in l_i$. Therefore, it follows due to Statement~3 that if  $(q_1',S')(x_i)$ occurs in $\xi$ then $s_i \in \text{dom}
		((q_1',S'))$.
		Thus,  $\p{(q_1,S)}^{\hat{T}_1}  (s'))\Rightarrow t_1$ where
		\[
		t_1\in \xi [(q_1',S')(x_i)\leftarrow \p{(q_1',S')}^{\hat{T}_1} (s_i) \mid (q_1',S')\in Q_1',i\in [k]] 
		\]
		and $\p{q_2}^{T_2} (t_1) \Rightarrow r_1' $ where
		\[
		r_1'\in \psi[q_2'(u) \leftarrow \p{q_2'}^{T_2} (t_1/u) \mid q_2'\in Q_2, u\in V(\xi)].
		\]
		Note that the tree $t_1/u$ is produced by the state $(q_1',S')$ on input $s_i$ if
		$\xi [u] = (q_1',S') (x_i)$. 
		We remark that the node $g$ is labeled by $q_2(u)$ in $\psi$ where $u$ is a node of  $\xi$ such that $u$ is labeled by a symbol of the form
		$(q_1',S') (x_i)$  if and only if $g$ is labeled by $(q_1',S',q_2') (x_i)$ in $\gamma$ due to Statement~2.
		Due to  $(*)$, it follows that  \[\p{(q_1',S',q_2')}^M (s_i)=\{r'\}=\p{q_2'}^{T_2} (t_1/u).\]
		Therefore, $(*)$ yields $r_1'/g=r_1/g$.
		Due to the definition of $\gamma$ and $\psi$, i.e. Statement~2, our claim follows.
	\end{claimproof}
\end{proof*}